\documentclass[12pt,a4paper]{article}
\usepackage{amssymb}
\usepackage{amsmath}
\usepackage{a4wide}
\usepackage{epsfig}

\numberwithin{equation}{section}
\newcommand{\ade}{\mbox{$A$-$D$-$E$ }}

\newcommand{\Bild}[4]{
\begin{figure}[htb]
  \begin{center}
    \leavevmode
    \epsfig{file=#2,height=#1cm}
    \caption{{\small #3}}
    \label{#4}
  \end{center}
\end{figure}}

\newcommand{\BWpq}[7]{ W^{#1,#2} \left( \left. \begin{array}{cc} #6 & #5
\\ #3 & #4
\end{array} \right| #7 \right) }

\newcommand{\BBpq}[6]{ W^{#1 #2} \left( \begin{array}{cc} #6 & #5
\\ #3 & #4
\end{array}  \right) }

\newcommand{\gsin}[2]{\left[ \begin{array}{c} #1 \\ #2 \end{array}\right]}
\newcommand{\psin}[2]{\left[ #1 \right]_{#2}}

\newcommand{\DR}[6]{ B^{#1}_{#2} \left( \left. #3 \begin{array}{c} #5\\ #4\end{array}
\right| #6 \right) }
\newcommand{\DL}[6]{ B^{#1}_{#2} \left( \left. \! \begin{array}{c} #4\\ #3 \end{array} #5
 \, \right| #6 \right) }

\newcommand{\BR}[5]{ B^{#1} \left( \left. #2 \begin{array}{ccc} #4 \\ 
#3 \end{array} \right| #5 \right) }

\newcommand{\BL}[5]{ B^{#1} \left( \left. \begin{array}{ccc} #3 \\ 
#2 \end{array} #4 \right| #5 \right) }

\def\pb{\raisebox{3ex}{}\raisebox{-2ex}{}}

\def\l{\lambda}
\def\bdm{\begin{displaymath}}
\def\edm{\end{displaymath}}
\def\bea{\begin{eqnarray}}
\def\eea{\end{eqnarray}}

\begin{document}

\title{\bf Integrable Lattice Realizations of $N=1$ Superconformal Boundary Conditions}

\date{ \today}

\author{ Christoph~Richard and Paul~A.~Pearce \\
Department of Mathematics and Statistics\\
University of Melbourne, Victoria 3010, Australia}

\maketitle

\begin{abstract}
We construct integrable boundary conditions for $\widehat{s\ell}(2)$ coset models with 
central charges $c=\frac{3}{2}-\frac{12}{m(m+2)}$ and $m=3,4,\ldots$ 
The associated cylinder partition functions are generating functions for the branching 
functions but these boundary conditions manifestly break the superconformal symmetry. 
We show that there are additional integrable boundary conditions, satisfying the boundary 
Yang-Baxter equation, which respect the superconformal symmetry and lead to generating 
functions for the superconformal characters in both Ramond and Neveu-Schwarz sectors. 
We also present general formulas for the cylinder partition functions.
This involves an alternative derivation of the superconformal Verlinde formula
recently proposed by Nepomechie.
\end{abstract}

\section{Introduction}

It is known that for certain families of rational conformal
field theories (CFTs)~\cite{BPPZ98, BPPZ00} it is possible to construct
complete sets of integrable and conformal boundary
conditions.  More specifically, if the associated critical
Yang-Baxter integrable lattice model is known, then fusion
techniques can be used to construct integrable boundary
conditions which satisfy the boundary Yang-Baxter equation and
give rise to all of the conformal boundary conditions in the
continuum scaling limit.  This program has been carried out in
particular for $\widehat{s\ell}(2)$ minimal~\cite{BP01} and ${\mathbb Z}_k$ parafermion
models~\cite{MP01}. 
In these cases the Virasoro characters and parafermionic string 
functions are dictated by the relevant chiral algebra. 
In some cases, however, there exists an extended chiral symmetry 
and in such situations, at least from the viewpoint of CFT, 
the actual chiral algebra which is used is a matter of choice 
depending on the symmetries which are to be preserved. 
A relevant question is then whether integrable and conformal 
boundary conditions can be obtained which are compatible with 
the extended chiral symmetry. 
If the answer is yes, as we expect is generally the case, 
then this observation necessarily implies the existence of 
new solutions to the boundary Yang-Baxter equations for the 
underlying critical lattice model.

In this paper we consider the level two $\widehat{s\ell}(2)$ coset models
which can alternatively be viewed as $N=1$ superconformal
theories. 
We review these theories from the different CFT viewpoints in 
Section~2. 
We give a generalized Verlinde formula for the fusion coefficients of the superconformal
theories, being valid for all values of the central charge.
This generalizes a formula recently proposed by Nepomechie \cite{N01} for the
theories without fixed point, who extended the analysis of \cite{AS98,AS99}, 
using a different approach than ours.
In Section~3 we define lattice realizations and use a generalized fusion 
procedure to construct integrable boundary conditions.
In the case corresponding to the usual fusion procedure, this leads 
to integrable and conformal boundary conditions for the coset models.
The analysis generalises straightforwardly to general coset theories.
We thus extend the results of \cite{AR99}, where only a subset of all coset 
boundary conditions had been found. 
The generalized fusion leads to additional integrable boundary conditions 
which we posit to be compatible with the superconformal symmetry. 
This is explained in Section~4.
For $m$ odd, we give boundary weights corresponding to all coset and superconformal
boundary fields.
For $m$ even, the construction is complete with exception of the fixed point fields.

In Section~5 we confirm numerically that these solutions 
of the boundary Yang-Baxter equations indeed lead to the branching
functions and to the superconformal characters in the continuum scaling limit.
The integrable superconformal boundary conditions can be extended off-criticality
and are highly relevant to the study of superconformal bulk and boundary flows via TBA 
\cite{PCA01}.

\section{Superconformal theories}

In this section we review the properties of $N=1$ superconformal theories, whose unitary
highest weight representations have central charges
\begin{equation}\label{form:charges}
c=\frac{3}{2}\left[  1- \frac{8}{m(m+2)}\right] \qquad m=3,4,\ldots
\end{equation}
We focus on the  $A$-type of the \ade classification \cite{C87} of torus partition 
functions.
We give an alternative description using the coset construction and relate both
approaches.
We give explicit expressions for the $S$ matrices and derive the fusion rules.
We derive a generalized Verlinde formula, which describes the fusion of superconformal
boundary conditions, following the general framework of Behrend, Pearce, Petkova and 
Zuber \cite{BPPZ98,BPPZ00}. 

\subsection{Coset construction}

The coset description of these models is given by the coset \cite{GKO86, KP89, SY90}
\begin{equation}\label{form:coset}
\frac{\widehat{s\ell}(2)_{m-2} \otimes \widehat{s\ell}(2)_2}{\widehat{s\ell}(2)_m}.
\end{equation}
Its branching functions $b_{r,s}^{(l)}(q)$ satisfy
\begin{equation}
\chi_{r-1,m-2}(q,z)\chi_{l,2}(q,z)=\sum_{s=1}^{m+1}b_{r,s}^{(l)}(q)\chi_{s-1,m}(q,z),
\end{equation}
where the ranges of the indices are $1\le r \le m-1$, $1\le s \le m+1$, $0\le l \le 2$.
The $\chi_{r,s}(q,z)$ are the characters of the affine Lie algebra 
$\widehat{s\ell}(2)$ at levels $m-2$, $2$ and $m$, respectively.
The branching functions satisfy
\begin{eqnarray}\label{form:symm}
b_{r,s}^{(l)}(q) &=& b_{m-r,m-s+2}^{(2-l)}(q), \\
b_{r,s}^{(l)}(q) &=& 0, \qquad r+s+l = 1 \mbox{ mod } 2. \nonumber 
\end{eqnarray}
The weights of the non-vanishing branching functions are given by
\begin{equation}
\begin{split}
\Delta_{r,s}^{(l)}= 
&\frac{[(m+2)r-ms]^2-4}{8m(m+2)} + \frac{1}{8}\left(
\frac{3}{4}-(-1)^{(l+s-r)/2} \right)\left(1+(-1)^{r-s}\right)\\
&+\frac{1}{16} +
\delta_{l,0}\delta_{r,m-1}\delta_{s,m+1}+\delta_{l,2}\delta_{r,1}\delta_{s,1}.
\end{split}
\end{equation} 
For $m=3$, we obtain the weight tables as shown below.
\bdm
\begin{array}{l|c|c|c|c|r}
\multicolumn{6}{l}{r \qquad l=0} \\
\cline{2-5}
 & \pb \hspace{2ex}  & \frac{1}{10} & \hspace{2ex} & \frac{3}{2} &\\
\cline{2-5}
& \pb 0 &  & \frac{3}{5} & \hspace{2ex}  &\\
\cline{2-5}
\multicolumn{6}{r}{s}
\end{array}
\qquad
\begin{array}{l|c|c|c|c|r}
\multicolumn{6}{l}{r \qquad \quad l=1} \\
\cline{2-5}
& \pb \frac{7}{16} &  & \frac{3}{80} & & \\
\cline{2-5}
& \pb  & \frac{3}{80} & & \frac{7}{16} &\\
\cline{2-5}
\multicolumn{6}{r}{s}
\end{array}
\qquad
\begin{array}{l|c|c|c|c|r}
\multicolumn{6}{l}{r \qquad l=2} \\
\cline{2-5}
& \pb \hspace{2ex} & \frac{3}{5} &  & 0 & \\
\cline{2-5}
& \pb \frac{3}{2} & \hspace{2ex} & \frac{1}{10} & \hspace{2ex} & \\
\cline{2-5}
\multicolumn{6}{r}{s}
\end{array}
\edm
For $m=4$, we obtain the following set of weights:
\bdm
\begin{array}{l|c|c|c|c|c|r}
\multicolumn{7}{l}{r \qquad \quad l=0} \\
\cline{2-6}
& \pb \frac{3}{2} &  & \frac{1}{6} & & \frac{3}{2} &\\
\cline{2-6}
& \pb \hspace{2ex} & \frac{1}{16} & \hspace{2ex}  &  \frac{9}{16} & \hspace{2ex} &\\
\cline{2-6}
& \pb 0 &  & \frac{2}{3} &  & 1 & \\
\cline{2-6}
\multicolumn{7}{r}{s}
\end{array}
\qquad
\begin{array}{l|c|c|c|c|c|r}
\multicolumn{7}{l}{r \qquad \quad l=1} \\
\cline{2-6}
& \pb \hspace{2ex} & \frac{9}{16} &  & \frac{1}{16} & \hspace{2ex} & \\
\cline{2-6}
& \pb \frac{3}{8} & & \frac{1}{24} &  &  \frac{3}{8} & \\
\cline{2-6}
& \pb  & \frac{1}{16} &  & \frac{9}{16} & &  \\
\cline{2-6}
\multicolumn{7}{r}{s}
\end{array}
\qquad
\begin{array}{l|c|c|c|c|c|r}
\multicolumn{7}{l}{r \qquad \quad l=2} \\
\cline{2-6}
& \pb 1 &  & \frac{2}{3} &  & 0 &\\
\cline{2-6}
& \pb \hspace{2ex}  & \frac{9}{16} & \hspace{2ex} &  \frac{1}{16} & \hspace{2ex} & \\
\cline{2-6}
& \pb \frac{3}{2} & & \frac{1}{6} & & \frac{3}{2} & \\
\cline{2-6}
\multicolumn{7}{r}{s}
\end{array}
\edm
The branching functions can be expressed in terms of the branching coefficients
$d_{j_1j_2j_3}(q)$ defined \cite{DJKMO87,DJKMO88} by
\begin{equation}
\begin{split}
d_{j_1j_2j_3}(q)=&q^{j_1^2/4m_1+j_2^2/4m_2-j_3^2/4m_3-1/8}Q(q)^{-3}
\sum_{\epsilon_1,\epsilon_2=\pm1} \left( \sum _{k,n_1,n_2}^{(1)}-
\sum _{k,n_1,n_2}^{(2)}\right)\\
&\times (-1)^{k+(\epsilon_1+\epsilon_2)/2}q^{k(k-1)/2+k(j_3+1)/2+\sum_{i=1}^2 [k
\epsilon_i(m_i n_i+j_i/2)+m_i n_i^2+j_in_i]},
\end{split}
\end{equation}
where
\begin{equation}
Q(q) = \prod_{n=1}^\infty (1-q^n), \qquad m_1=m, \qquad m_2=4, \qquad m_3=m+2,
\end{equation}
and the two sums are restricted to values of $k$, $n_1$, $n_2$ satisfying
\begin{eqnarray}
\sum^{(1)} &:& k \ge \xi+1,\qquad \eta \le \frac{j_3+1}{2}+\sum_{i=1}^2 \epsilon_i \left(
m_i n_i + \frac{j_i}{2} \right) \in \mathbb{Z}, \\
\sum^{(2)} &:& k \le \xi,\qquad \eta-1 \ge \frac{j_3+1}{2}+\sum_{i=1}^2 \epsilon_i \left(
m_i n_i + \frac{j_i}{2} \right) \in \mathbb{Z}. \nonumber
\end{eqnarray}
The integers $\xi=\xi(\epsilon_i,n_i)$ and $\eta=\eta(\epsilon_i,n_i)$ can be chosen
arbitrarily for fixed $\epsilon_i$ and $n_i$.
The nonvanishing branching functions are given in terms of these as
\begin{equation}
b_{r,s}^{(l)}(q) = d_{r,l+1,s}(q).
\end{equation}
Under modular transformations, the branching functions transform as
\begin{equation}
b_{r,s}^{(l)}\left( e^{2\pi i \tau}\right) =
\sum_{r'=1}^{m-1}\sum_{s'=1}^{m+1}\sum_{l'=0}^2 
S_{(r,s,l)}^{(r',s',l')} \; b_{r',s'}^{(l')}
\left( e^{-2\pi i/ \tau} \right),
\end{equation}
where the transformation matrix $S$ satisfies
\begin{equation}
S^T = S^{-1}, \qquad S^2=\mathbb{I}.
\end{equation}
The entries of $S$ are given explicitly by
\begin{equation}
S_{(r,s,l)}^{(r',s',l')} = \frac{\sqrt 2}{\sqrt{m(m+2)}} \sin\frac{\pi r r'}{m} 
\sin\frac{\pi s s'}{m+2} \sin\frac{\pi (l+1) (l'+1)}{4}.
\end{equation}
Note that the $S$ matrix above cannot be interpreted as the modular matrix of an
appropriate conformal field theory since its definition includes vanishing branching
functions, which do not correspond to primary fields.
A method how to resolve this problem is given in \cite{SY90}.
For completeness, we will discuss their method in the following paragraph.

We conclude this subsection with a discussion of the Verlinde coefficients
\begin{equation}\label{form:cosetVer}
n_{(r,s,l),(r_1,s_1,l_1)}^{(r_2,s_2,l_2)} =
\sum_{r'=1}^{m-1} \sum_{s'=1}^{m+1} \sum_{l'=0}^2 \frac{
S_{(r,s,l)}^{(r',s',l')}
S_{(r_1,s_1,l_1)}^{(r',s',l')}
(S^{-1})_{(r_2,s_2,l_2)}^{(r',s',l')}}{S_{(1,1,0)}^{(r',s',l')}}
\in \mathbb{N}_0.
\end{equation}
As argued above, these coefficients cannot be interpreted as fusion
coefficients for fusion of primary operators of an appropriate
conformal field theory.
They are, however, closely related to the fusion coefficients of a modified
description, which is relevant to our discussion of the related lattice models.
Due to the coset construction, the Verlinde coefficients can
be written in tensor product form
\begin{equation}
n_{(r,s,l),(r',s',l')}^{(r'',s'',l'')} = n_{r r'}^{(m)\, r''} 
n_{s s'}^{(m+2)\, s''} n_{l+1, l'+1}^{(4)\, l''+1},
\end{equation}
where $n_{ij}^{(g)\,k}$ are the fusion coefficients of the affine Lie algebra 
$\widehat{s\ell}(2)$ at level $g-2$.
The fusion coefficients $n_{ij}^{(g)\,k}$ can be expressed in terms of the matrix 
elements of the fused adjacency matrices $F^{(g)\,r}$ of the graph $A_{g-1}$
as
\begin{equation}\label{form:fusmat}
n_{ij}^{(g)\,k}=F_{i,j}^{(g)\,k},
\end{equation}
where $F^{(g)\,r}$ are given recursively in terms of the adjacency matrix of the 
graph $A_{g-1}$  by the $s\ell(2)$ fusion rules
\begin{equation}
F^{(g)\,r}=A_{g-1} F^{(g)\,r-1}-F^{(g)\,r-2}, \qquad r=3,\ldots,g-1
\end{equation} 
with initial conditions
\begin{equation}
F^{(g)\,1}=\mathbb{I}_{g-1}, \qquad F^{(g)\,2}=A_{g-1}.
\end{equation} 

\subsection{Field identification and fixed point resolution}

The method proposed in \cite{SY90} consists of two steps: field identifaction and fixed
point resolution.
We first define a fundamantal domain $E=E_0 \cup E_1 \cup E_f \cup E_2$ of $(r,s,l)$ values:
\begin{eqnarray}
E_{0} &=& \left\{ (r,s,0) \, | \,  r-s \mbox{ mod } 4=0  \right\}, \\
E_{2} &=& \left\{ (r,s,2) \, | \,  r-s \mbox{ mod } 4=0  \right\}, \nonumber\\
E_{1} &=& \left\{ (r,s,1) \, | \,  r-s \mbox{ mod } 2= 1, s \le (m+1)/2; \right.
\nonumber \\
& & \left.  (-1)^{r-s} = \pm 1, s= m/2+1, r < m/2 \right\}, \nonumber \\
E_{f} &=& \left\{ (m/2,m/2+1,1), m \mbox{ even} \right\}. \nonumber
\end{eqnarray}
The second step consists in resolving fixed points under the transformation
(\ref{form:symm}).
A fixed point branching function occurs for $m$ even and has labels 
$(r,s,l)=(m/2,m/2+1,1)$.
The resolution is done by duplicating the fixed point branching function
\cite{SY90} according to 
\begin{equation}\label{form:brfpfield}
b_f(q) := b^{(1)}_{m/2,m/2+1}(q) \mapsto 
\left\{
\begin{array}{c}
b_{f_1}(q):=\frac{1}{2} \left( b_f(q) - 1 \right),\\ 
b_{f_2}(q):=\frac{1}{2} \left( b_f(q) + 1 \right).
\end{array}
\right.
\end{equation}
In this new basis, the resulting modular matrix $\widetilde{S}$ is given by
\begin{equation}\label{form:Sb}
\widetilde{S}=\left( \begin{array}{ccc}
2 S_{a,b} & S_{a,f} & S_{a,f} \\
S_{f,b} & \frac{1}{2} &  - \frac{1}{2}\\
S_{f,b} & -\frac{1}{2} &  \frac{1}{2}
\end{array} \right),
\end{equation}
where $a,b \in E_0\cup E_1 \cup E_2$ and $f\in E_f$.
For $m$ odd, the matrix $\widetilde{S}$ is simply
\begin{equation}\label{form:Sb2}
\widetilde{S}=\left( \begin{array}{c}
2 S_{a,b} 
\end{array} \right),
\end{equation}
where $a,b \in E_0\cup E_1\cup E_2$.
It satisfies in both cases
\begin{equation}
\widetilde{S}^T = \widetilde{S}^{-1}, \qquad \widetilde{S}^2=\mathbb{I}.
\end{equation}
Fusion coefficients for the corresponding primary fields $\Phi_i$ can
be computed using the Verlinde formula for $\widetilde{S}$,
\begin{equation}\label{form:cosVer}
n_{ij}^k = \sum_n \frac{\widetilde{S}_{in}\widetilde{S}_{jn}
\widetilde{S}^{-1}_{nk}}{\widetilde{S}_{1n}}.
\end{equation}
The labels $i,j,k,n$ denote the values $(r,s,l) \in E_0 \cup E_1\cup E_2$ 
resp. $f_1,f_2$ of the corresponding primary fields.
The label $1$ corresponds to the vacuum field.
It can be checked that the fusion coefficients are integers.
They coincide with the Verlinde coefficients (\ref{form:cosetVer}),
restricted to the above fundamental domain,
after a change of basis for the fixed point branching fields to its
sum and difference according to (\ref{form:brfpfield}).

\subsection{Superconformal data}

The unitary heighest weight representations of the $N=1$ superconformal algebra
have central charge (\ref{form:charges}) and conformal dimensions
\begin{equation}
\Delta_{r,s}= 
\frac{[(m+2)r-m s]^2-4}{8m(m+2)} +\frac{1}{32}\left[1-(-1)^{r-s}\right],
\end{equation}
where $1\le r\le m-1$ and $1\le s\le m+1$.
The cases $r-s$ even or odd correspond to the Neveu-Schwarz and to the Ramond sector,
respectively.
For $m=3$ and $m=4$, the Kac table of conformal dimensions are
\bdm
\begin{array}{l|c|c|c|c|r}
\multicolumn{6}{l}{r \qquad m=3} \\
\cline{2-5}
& \pb \frac{7}{16} & \frac{1}{10} & \frac{3}{80} & 0 & \\
\cline{2-5}
& \pb 0 & \frac{3}{80} & \frac{1}{10} & \frac{7}{16} & \\
\cline{2-5}
\multicolumn{6}{r}{s}
\end{array}
\qquad
\begin{array}{l|c|c|c|c|c|r}
\multicolumn{7}{l}{r \qquad \quad m=4} \\
\cline{2-6}
& \pb 1 & \frac{9}{16} & \frac{1}{6} &  \frac{1}{16} & 0 &\\
\cline{2-6}
& \pb \frac{3}{8} & \frac{1}{16} & \frac{1}{24} &  \frac{1}{16} & \frac{3}{8} &\\
\cline{2-6}
& \pb 0 & \frac{1}{16} & \frac{1}{6} & \frac{9}{16} & 1 & \\
\cline{2-6}
\multicolumn{7}{r}{s}
\end{array}
\edm
Note that these tables may be obtained by combining the appropriate coset tables.
In the Neveu Schwarz sector $r-s$ even, this amounts to identifying fields corresponding
to superpartners.\\
The superconformal characters are given by \cite{MY86,C87}
\begin{eqnarray}\label{form:scchar}
\chi_{r,s}^{NS}(q) &=& q^{-c/24} \prod_{n=1}^\infty\frac{1+q^{n-1/2}}{1-q^n}
\sum_{n=-\infty}^\infty \left( q^{\gamma_{r,s}(n)} - q^{\gamma_{-r,s}(n)}   \right) \\
\chi_{r,s}^{\widetilde{NS}}(q) &=& q^{-c/24} \prod_{n=1}^\infty\frac{1-q^{n-1/2}}{1-q^n}
\sum_{n=-\infty}^\infty  (-1)^{mn} \left[ q^{\gamma_{r,s}(n)} - (-1)^{rs} 
q^{\gamma_{-r,s}(n)}   \right] \nonumber \\
\chi_{r,s}^{R}(q) &=&
q^{-c/24+ 1/16}\prod_{n=1}^\infty\frac{1+q^n}{1-q^n}\sum_{n=-\infty}^\infty
\left( q^{\gamma_{r,s}(n)} - q^{\gamma_{-r,s}(n)}   \right) \nonumber \\
\chi_{r,s}^{\widetilde R} &=& \delta_{(r,s),(m/2,m/2+1)},\nonumber
\end{eqnarray}
where
\begin{equation}
\gamma_{r,s}(n)=\frac{\left[2m(m+2)n-r(m+2)+sm\right]^2-4}{8m(m+2)}.
\end{equation}
The trivial character $\chi_{r,s}^{\widetilde R}$ in the sector $\widetilde{R}$ occurs
for $m$ even only.
The $S$ matrix for the superconformal theories is usually defined in terms of modified
characters in the Ramond sector, since the $R$ states are doubly degenerate except for
the fixed point $(r,s)=(m/2,m/2+1)$.
We thus define modified characters 
${\widehat \chi}_{r,s}^{R}(q)= g_{r,s}\sqrt{2} \chi_{r,s}^{R}(q)$,
where
\begin{equation}
g_{rs} = \left\{ 
\begin{array}{cl} 
1 &  (r,s) \neq (m/2,m/2+1), \\ 
\frac{1}{\sqrt{2}} & \mbox{otherwise.} 
\end{array} 
\right.
\end{equation}
For the definition of the superconformal $S$-matrix, we restrict the values of the
conformal labels $(r,s)$ to the following fundamental domain
\begin{eqnarray}\label{form:fdom}
E_{\widetilde{NS}}=E_{NS} &=& \left\{ (r,s) \, | \, r-s \mod 4 =0 \right\}, \\
E_{R} &=& \left\{ (r,s) \, | \,   r-s \mbox{ odd}, s\le (m+1)/2; \right. \nonumber \\
&& \left. r-s \mbox{ odd}, s = m/2+1, r\le m/2 \right\}. \nonumber
\end{eqnarray}
This is equivalent to the choice in \cite{MY86}.
In this modified basis and for $m$ odd, the $S$ matrix is given by \cite{MY86}
\begin{equation}\label{form:Schi1}
S = \left( 
\begin{array}{ccc} 
S^{[NS,NS]} & 0 & 0 \\
0 & 0 & S^{[\widetilde{NS},R]}\\
0 & S^{[R,\widetilde{NS}]} & 0
\end{array} \right),
\end{equation}
with matrix elements
\begin{eqnarray}
S^{[NS,NS]}_{(rs),(r's')} &=& 
\frac{4}{\sqrt{m(m+2)}}\sin\frac{\pi r r'}{m} \sin\frac{\pi s s'}{m+2}, \\
S^{[\widetilde{NS},R]}_{(rs),(r's')} &=& 
(-1)^{(r-s)/2}\frac{4 g_{r's'}}{\sqrt{m(m+2)}}\sin\frac{\pi r r'}{m} 
\sin\frac{\pi s s'}{m+2},\nonumber\\
S^{[R,\widetilde{NS}]}_{(rs),(r's')} &=&
(-1)^{(r'-s')/2}\frac{4 g_{r,s}}{\sqrt{m(m+2)}}\sin\frac{\pi r r'}{m} 
\sin\frac{\pi s s'}{m+2}. \nonumber
\end{eqnarray}
For $m$ even, the $S$ matrix is given by \cite{MY86}
\begin{equation}
S = \left( 
\begin{array}{cccc}\label{form:Schi2}
S^{[NS,NS]} & 0 & 0 & 0 \\
0 & 0 & S^{[\widetilde{NS},R]} & 0\\
0 & S^{[R,\widetilde{NS}]} & 0 & 0\\
0 & 0 & 0 & 1
\end{array} \right),
\end{equation}
where the last entry corresponds to the trivial character in the sector $\widetilde{R}$.
In both cases, the $S$ matrix satisfies
\begin{equation}
S^T = S^{-1}, \qquad S^2=\mathbb{I}.
\end{equation}
For $m$ odd, a generalized Verlinde formula for the fusion coefficients has
been given in \cite{N01}.
We give a generalization of this formula comprising both cases $m$ odd
and $m$ even, which will be deduced below.
Non-vanishing fusion coefficients occur in the sectors
\begin{equation}
\begin{split}
&NS \times  NS \to NS, \qquad 
\widetilde{NS} \times \widetilde{NS} \to \widetilde{NS},\\
&NS \times R \to R, \qquad \quad
\widetilde{NS} \times \widetilde{R} \to \widetilde{R},\\
&R \times R \to NS, \qquad \quad
\widetilde{R} \times \widetilde{R} \to \widetilde{NS}.
\end{split}
\end{equation}
They are given explicitly by
\begin{eqnarray}\label{form:genVerlinde}
n_{{NS}_i,{NS}_j}^{{NS}_k} &=& \sum_{l\in E_{NS}} 
\frac{S^{[NS,NS]}_{il}S^{[NS,NS]}_{jl}
(S^{[NS,NS]})_{lk}^{-1}}{S^{[NS,NS]}_{1l}},\\
n_{\widetilde{NS}_i,\widetilde{NS}_j}^{\widetilde{NS}_k} &=& \sum_{l\in E_{R}} 
\frac{S^{[\widetilde{NS},R]}_{il}S^{[\widetilde{NS},R]}_{jl}
(S^{[\widetilde{NS},R]})_{lk}^{-1}}
{S^{[\widetilde{NS},R]}_{1l}}, \nonumber\\
n_{R_i,R_j}^{{NS}_k} &=& \frac{1}{g_i g_j} \sum_{l\in E_{NS}} 
\frac{S^{[R,\widetilde{NS}]}_{il}S^{[R,\widetilde{NS}]}_{jl}(S^{[NS,NS]})_{lk}^{-1}}
{S^{[NS,NS]}_{1l}}, \nonumber\\
n_{{NS}_i,R_j}^{R_k} &=& \frac{g_k}{g_j}\sum_{l\in E_{NS}} 
\frac{S^{[NS,NS]}_{il}S^{[R,\widetilde{NS}]}_{jl}(S^{[R,\widetilde{NS}]})_{lk}^{-1}}
{S^{[NS,NS]}_{1l}}, \nonumber
\end{eqnarray}
\begin{equation*}
n_{\widetilde{R},\widetilde{R}}^{\widetilde{NS}_k} = 
2 \, n_{\widetilde{NS}_k,\widetilde{R}}^{\widetilde{R}} =
2 \frac{S^{[\widetilde{NS},R]}_{kf}}{S^{[\widetilde{NS},R]}_{1f}}.
\end{equation*}
These expressions coincide with \cite{N01} up to factors of two, which
arise from different normalizations.
We adopted the convention that fusion of the vacuum with Ramond states
yields the Ramond state back, wheras in \cite{N01}, eqn.~(3.17), twice the Ramond
state is obtained. 
It can be checked that the fusion coefficients are integers.
We emphasize that these formulae can be obtained from the fusion coefficients of the 
coset construction by performing a change of basis to superconformal primary fields,
as explained in following paragraph.

\subsection{Branching functions and superconformal characters}

The branching functions are related to the superconformal characters by a linear
transformation $M$
\begin{equation}\label{form:trafo}
\chi=M\,b, \qquad S_\chi=M\,S_b\,M^{-1},
\end{equation}
which also relates the two $S$ matrices (\ref{form:Sb}) and (\ref{form:Schi1}) resp. 
(\ref{form:Schi2}).
The branching coefficients are expressed in terms of the (modified)
superconformal characters by
\begin{eqnarray} \label{form:braco}
b_{r,s}^{(0)}(q) &=& \frac{1}{2}\left[ \chi_{r,s}^{NS}(q)+  
(-1)^{\frac{r-s}{2}}\chi_{r,s}^{\widetilde{NS}}(q)
\right], \qquad  r-s \mbox{ even}, \\ 
b_{r,s}^{(2)}(q) &=& \frac{1}{2}\left[ \chi_{r,s}^{NS}(q)- 
(-1)^{\frac{r-s}{2}}\chi_{r,s}^{\widetilde{NS}}(q)
\right], \qquad  r-s \mbox{ even}, \nonumber\\
b_{r,s}^{(1)}(q) &=&  \frac{1}{\sqrt{2}}{\widehat \chi}_{r,s}^{R}(q), 
\qquad  (r,s) \neq(m/2,m/2+1)  
\qquad  r-s \mbox{ odd}, \nonumber\\
b_{f_1}(q) &=& \frac{1}{2} \left[ 
{\widehat \chi}_f^R(q) - \chi_f^{\widetilde R} \right],
\qquad f= (m/2,m/2+1), \nonumber\\
b_{f_2}(q) &=& \frac{1}{2} \left[ 
{\widehat \chi}_f^R(q) + \chi_f^{\widetilde R} \right],
\qquad f= (m/2,m/2+1).
\end{eqnarray}
This relation is invertible and gives the (modified) superconformal
characters in terms of the branching functions as
\begin{eqnarray}\label{form:Mscf}
\chi_{r,s}^{NS}(q) & = & b_{r,s}^{(0)}(q) + b_{r,s}^{(2)}(q), 
\qquad  r-s \mbox{ even}, \\ 
\chi_{r,s}^{\widetilde{NS}}(q) &=& (-1)^{(r-s)/2} 
\left( b_{r,s}^{(0)}(q) - b_{r,s}^{(2)}(q) \right), \qquad  r-s \mbox{ even}, \nonumber\\
{\widehat \chi}_{r,s}^{R}(q) &=& \sqrt{2} \, b_{r,s}^{(1)}(q),  
\qquad  (r,s) \neq(m/2,m/2+1), \qquad  r-s \mbox{ odd}, \nonumber\\
{\widehat \chi}_f^R(q) &=& b_{f_1}(q) + b_{f_2}(q), \qquad f= (m/2,m/2+1), \nonumber\\
\chi_f^{\widetilde R} &=& b_{f_1}(q) - b_{f_2}(q), \qquad f= (m/2,m/2+1). \nonumber
\end{eqnarray}

We now discuss the relation between the fusion coefficients of the coset
description and the superconformal description.
This leads to an alternative derivation of the generalized Verlinde formula, which is
valid also for the case $m$ even.
Under the change of basis of the primary fields
\begin{equation}
\widetilde{\Phi}=M\,\Phi, \qquad \widetilde{S}=M\,S\,M^{-1},
\end{equation}
fusion coefficients are given by a 
modified Verlinde formula
\begin{equation}\label{form:genVer}
\widetilde{n}_{ij}^k = \sum_n \frac{(\widetilde{S}M)_{in}(\widetilde{S}M)_{jn}
(\widetilde{S}M)^{-1}_{nk}}{(M^{-1}\widetilde{S}M)_{1n}}.
\end{equation}
The primary coset fields and superconformal fields are related via
\begin{eqnarray}
\Phi_{r,s}^{NS} & = & (\Phi_{r,s}^{(0)} + \Phi_{r,s}^{(2)})/2, 
\qquad  r-s \mbox{ even}, \\ 
\Phi_{r,s}^{\widetilde{NS}} &=& (-1)^{(r-s)/2} 
\left( \Phi_{r,s}^{(0)} - \Phi_{r,s}^{(2)} \right)/2, 
\qquad  r-s \mbox{ even}, \nonumber \\
\Phi_{r,s}^{R} &=& \frac{1}{\sqrt{2}}\, \Phi_{r,s}^{(1)}, 
\qquad  (r,s) \neq(m/2,m/2+1),
\qquad  r-s \mbox{ odd}, \nonumber\\
\Phi_f^R &=& \frac{1}{\sqrt{2}}\,(\Phi_{f_1}+\Phi_{f_2}), 
\qquad f = (m/2,m/2+1), \nonumber \\
\Phi_f^{\widetilde{R}} &=&
\frac{1}{\sqrt{2}}\,(\Phi_{f_1}-\Phi_{f_2}), 
\qquad f = (m/2,m/2+1). \nonumber 
\end{eqnarray}
Note that the matrix $M$ of basis change is different from the corresponding one for the
branching functions and superconformal characters (\ref{form:Mscf}).
It can be checked that the fusion coefficients for the superconformal primary fields are
integers, and that they are related to the $S$ matrix of the superconformal characters
(\ref{form:Schi1}) and (\ref{form:Schi2}) by the formulae 
(\ref{form:genVerlinde}) given above.

\subsection{Cylinder partition functions}

The modular invariant partition functions of the coset models (\ref{form:coset}) 
have been classified by Cappelli \cite{C87} in terms of a pair of graphs $(G',G)$ 
where $G'$ is of $A$-type or $D$-type, and $G$ is of $\ade$ type.
Throughout the paper, we restrict ourselves to $A$-type models, whose allowed spin
values are given by the adjacency matrix of the graph $A$. 

According to \cite{C87}, the modular invariant torus partition function for the
$(A_{m-1},A_{m+1})$ models is given by
\begin{equation}
Z(q)= a \sum_{r-s\, even} \left( |\chi_{r,s}^{NS}(q) |^2 +  
|\widetilde\chi_{r,s}^{NS}(q) |^2
\right) \; 
+ a \sum_{r-s\, odd} |\sqrt{2}\chi_{r,s}^{R}(q) |^2 + 
b \, |\chi_f^{\widetilde R}|^2.
\end{equation}
The summation is over all allowed values of $r=1,\ldots,m-1$, $s=1,\ldots,m+1$, the
constants $a$ and $b$ are not specified by modular invariance.
The last term is a constant and occurs for $m$ even only.
The above modular invariant partition function may be expressed in terms of the
branching functions using (\ref{form:Mscf}) as
\begin{eqnarray}\label{form:torcoset}
Z(q) &=& 2a \sum_{r=1}^{m-1} \sum_{s=1}^{m+1} \sum_{l=0}^2 |b_{rs}^{(l)}(q)|^2 +b\\
&=& 2a \! \! \! \! \! \!  \sum_{(r,s)\neq (\frac{m}{2},\frac{m+2}{2})} 
\! \! \! \! \! \!  |b_{rs}^{(l)}(q)|^2 
+(2a+b)(b_{f_1}^2(q)+b_{f_2}^2(q)) + (4a-2b) b_{f_1}(q) b_{f_2}(q). \nonumber 
\end{eqnarray}
In the above formula, the resolved fixed point branching function occurs for $m$ even
only. 
In analogy to the reasoning in \cite{BPPZ98,BPPZ00}, we claim that the cylinder partition 
functions for the superconformal theories are given by
\begin{equation}\label{form:supercyl}
Z_{i|j}^{(sc)}(q) = 
\sum_k  n_{i,j}^{k} \chi_k(q).
\end{equation}
The fusion coefficients $n_{i,j}^{k}$ are given by the generalized Verlinde formula
(\ref{form:genVerlinde}), and $\chi_k(q)$ are the superconformal
characters (\ref{form:scchar}).
The indices $i,j,k$ refer to the fundamental domain where the $S$
matrices (\ref{form:Schi1}) and (\ref{form:Schi2}) are defined.
For the coset description, we claim that cylinder partition functions are given by
\begin{equation}\label{form:coset2cyl}
Z_{i|j}^{(br)}(q) = 
\sum_k  n_{i,j}^{k} b_k(q),
\end{equation}
where the fusion coefficients $n_{i,k}^k$ are now given by the Verlinde formula
(\ref{form:cosVer}), the summation ranging over the fundamental domain
of the $S$ matrices (\ref{form:Sb2}) and (\ref{form:Sb}).

In the following, we will define integrable lattice models with a boundary, which
provide lattice realizations of the coset boundary conditions and superconformal 
boundary conditions as discussed above.
For each boundary field, we give an integrable boundary weight, with one subtlety for
the fixed point boundary conditions:
From our construction, we only get a lattice realization of the fixed point 
character $\chi_f^R(q)=b_f(q)$, but not of the trivial fixed point character in the
sector ${\widetilde R}$. 

\section{Lattice realization}

In this section we discuss lattice realizations of the coset and superconformal 
theories on a cylinder.
We cite explicit expressions for the face weights of $A$-type lattice models
\cite{DJMO86} at arbitrary fusion level $(p,q)$.
We explain how to construct integrable boundary weights using the fusion principle
and define double-row transfer matrices, generalizing the methods in 
\cite{BPO95,BP01}.
This is then specified to the case $p=q=2$, which corresponds to the superconformal
theories.
We explain how the conformal data connect to the eigenvalues of the double-row transfer
matrices.

\subsection{Face weights and boundary weights}

We consider the critical $A$-type lattice models at fusion level $(p,q)$.
We denote the face weights by
\setlength{\unitlength}{1cm}
\begin{equation}
\BWpq{p}{q}{a}{b}{c}{d}{u}=
% face weight upright
\begin{picture}(2.2,1)
\put(0.6,-0.4){\framebox(1,1){$u$}}
\put(0.3,-0.7){$a$}
\put(1.7,-0.7){$b$}
\put(1.7,0.7){$c$}
\put(0.3,0.7){$d$}
\end{picture}
=
% diagonally oriented face weight 
\begin{picture}(2.2,1)
\put(0.4,0.1){\line(1,1){0.7}}
\put(0.4,0.1){\line(1,-1){0.7}}
\put(1.81,0.1){\line(-1,1){0.7}}
\put(1.81,0.1){\line(-1,-1){0.7}}
\put(1,0){$u$}
\put(0.1,0){$a$}
\put(1,-1){$b$}
\put(1.91,0){$c$}
\put(1,0.9){$d$}
\end{picture}
% adjust lower end
\raisebox{-\unitlength}{}
\end{equation}

The values of adjacent spins are constrained by the fused adjacency conditions. 
Specifically, nonzero  weights only occur for spins $a,b,c,d$ satisfying the 
adjacency condition $F^p_{ab}F^q_{bc}F^p_{cd}F^q_{da}=1$, 
where the adjacency matrices $F^r$ are defined in (\ref{form:fusmat}).
For $A_4$ and $A_5$ and fusion level $(p,q)=(2,2)$, which corresponds to the
superconformal theories, we obtain the fused adjacency diagrams as shown in Fig.
\ref{fig:fused}. 
\Bild{4}{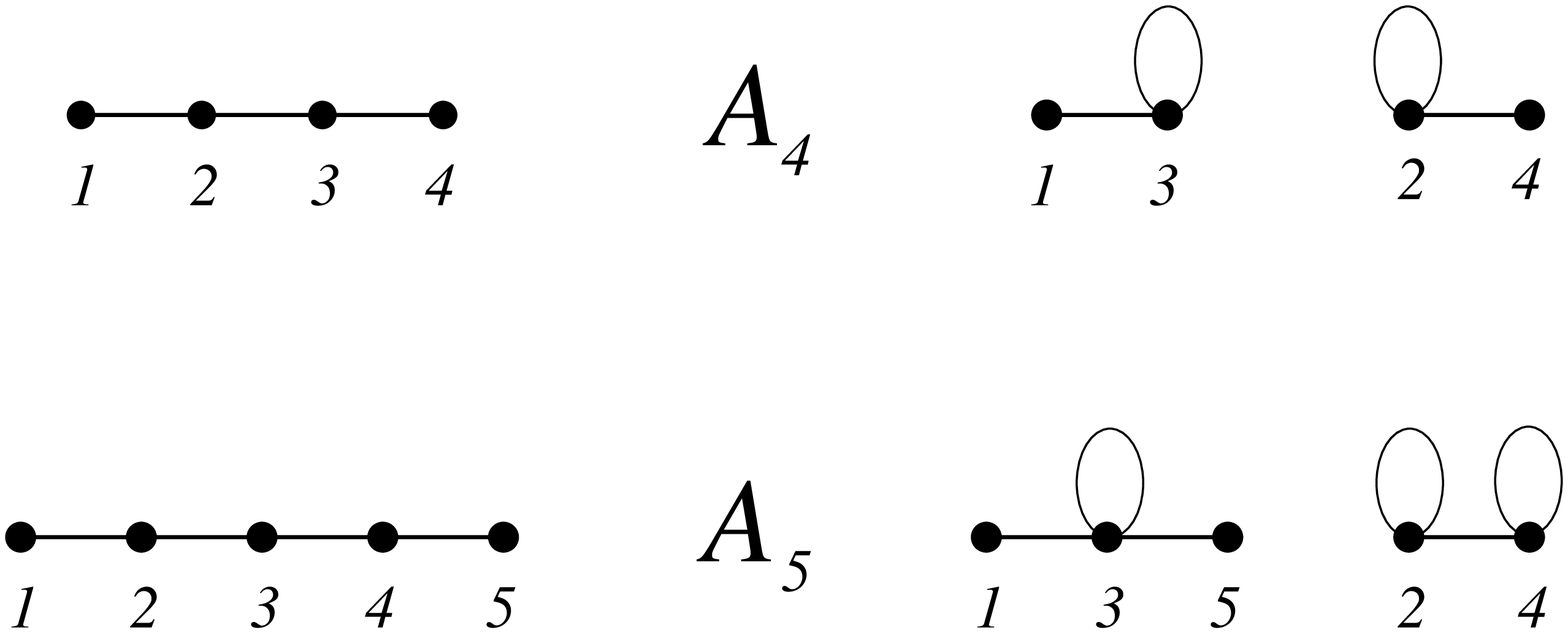}{Level two fused adjacency graphs for $A_4$ and $A_5$}{fig:fused}
At fusion level two, the fused adjacency diagrams consist of two separate parts,
corresponding to two disconnected lattice models.
For $A_L$ with $L$ odd, this leads to two identical, decoupled lattice models, wheras
the case $L$ even leads to different models, as we will see below.
We introduce the notation
\begin{equation}
\psin{u}{m} = \prod_{k=0}^{m-1}\sin(u-k\l), \qquad    
\gsin{u}{m} = \prod_{k=0}^{m-1}\frac{\sin(u-k\l)}{\sin(m\l-k\l)}.
\end{equation}
The non-vanishing weights have been given in explicit form in \cite{DJMO86}.
In either of the four cases $|a-b|=p$ or $|b-c|=q$ or $|c-d|=p$ or $|d-a|=q$ they have
the factorized form
\begin{equation}
\begin{split}
&\BWpq{p}{q}{a+2r-q}{a+2s-p+q}{a+2s-p}{a}{u} =\\
&\frac{\gsin{(p-s)\l}{q-r}\gsin{(a-s+r-p-1)\l}{r}\gsin{s\l-u}{r}\gsin{(a+s)
\l-u}{q-r}}{\gsin{(a+r)\l}{q-r}\gsin{(a+2r-q-1)\l}{r}}
\end{split}
\end{equation}
\begin{equation}
\begin{split}
&\BWpq{p}{q}{a+2r-q}{a+2s-p-q}{a+2s-p}{a}{u} =\\
&\frac{\gsin{s\l}{r}\gsin{(a+s)\l}{q-r}\gsin{(p-s)\l-u}{q-r}\gsin{(a+s-p+r-1)
\l+u}{r}}{\gsin{(a+r)\l}{q-r}\gsin{(a+2r-q-1)\l}{r}}
\end{split}
\end{equation}
Here, $\l=\pi/g$ is the spectral parameter, and $g$ is the Coxeter number of $A_{g-1}$.
Using these weights, the remaining ones are given by
\begin{equation}
\begin{split}
&\BWpq{p}{q}{a+2r-q}{b+2s-q}{b}{a}{u}\gsin{q\l}{s} =\\
&\sum_{j=\max(0,r+s-q)}^{\min(r,s)} \! \! \! \! \! \! \! \!
\BWpq{p}{s}{a+2j-s}{b+s}{b}{a}{u+(q-s)\l}
\BWpq{p}{q-s}{a+2r-q}{b+2s-q}{b+s}{a+2j-s}{u}
\end{split}
\end{equation}
These face weights satisfy the fused Yang-Baxter equation
\begin{equation}\label{form:YBE}
\begin{split}
\sum_g &\BWpq{r}{q}{a}{b}{g}{f}{u-v} \BWpq{p}{q}{b}{c}{d}{g}{u}
\BWpq{p}{r}{g}{d}{e}{f}{v} \\
& =\sum_g \BWpq{p}{r}{b}{c}{g}{a}{v} \BWpq{p}{q}{a}{g}{e}{f}{u}
\BWpq{r}{q}{g}{c}{d}{e}{u-v}
\end{split}
\end{equation}
This can be expressed graphically by
\begin{equation}
% diagonally oriented face weight 
\begin{picture}(4,1.5)(0,-0.1)
\put(-0.3,-0.1){$a$}
\put(0.9,-1.4){$b$}
\put(1.9,-1.4){$b$}
\put(3.6,-1.3){$c$}
\put(3.6,-0.1){$d$}
\put(3.6,1.0){$e$}
\put(0.9,1.1){$f$}
\put(1.9,1.1){$f$}
\put(2,0){\circle*{0.15}}
\put(0,0){\line(1,1){1}}
\put(0,0){\line(1,-1){1}}
\put(2,0){\line(-1,1){1}}
\put(2,0){\line(-1,-1){1}}
\put(2.1,-0.3){$g$}
\put(0.5,-0.1){$u-v$}
\put(2,0){\framebox(1.5,1){$v$}}
\put(2,-1){\framebox(1.5,1){$u$}}
\multiput(1,1)(0.09,0){12}{\circle*{0.02}}
\multiput(1,-1)(0.09,0){12}{\circle*{0.02}}
\end{picture}
=
% diagonally oriented face weight 
\begin{picture}(4,1.5)(1.5,-0.1)
\put(1.7,-0.1){$a$}
\put(1.9,-1.4){$b$}
\put(3.4,-1.3){$c$}
\put(4.4,-1.3){$c$}
\put(3.2,-0.3){$g$}
\put(3.4,1.1){$e$}
\put(4.4,1.1){$e$}
\put(1.9,1.1){$f$}
\put(5.6,-0.1){$d$}
\put(3.5,0){\circle*{0.15}}
\put(3.5,0){\line(1,1){1}}
\put(3.5,0){\line(1,-1){1}}
\put(5.5,0){\line(-1,1){1}}
\put(5.5,0){\line(-1,-1){1}}
\put(4,-0.1){$u-v$}
\put(2,0){\framebox(1.5,1){$u$}}
\put(2,-1){\framebox(1.5,1){$v$}}
\multiput(3.5,1)(0.09,0){12}{\circle*{0.02}}
\multiput(3.5,-1)(0.09,0){12}{\circle*{0.02}}
\end{picture}
% adjust lower end
\raisebox{-1.5\unitlength}{}
\end{equation}
The face weights satisfy the fused inversion relation
\begin{equation}\label{form:inv}
\sum_e \BWpq{q}{r}{a}{b}{e}{d}{u} \BWpq{r}{q}{e}{b}{c}{d}{-u} =
\rho^{qr}(u) \rho^{rq}(-u) \delta_{ac}F^q_{ab} F^r_{ad},
\end{equation}
where $\rho^{qr}(u)$ are model dependent functions.
For us, only the case $p=q=2$ will be of interest.
In this case, the functions can be disregarded because they are common factors.
This will simplify the reflection equations, see below.
This can be graphically described by
\begin{equation}
\begin{picture}(4.5,1.5)
% diagonally oriented face weight 
\put(0,0){
% diagonally oriented face weight 
\begin{picture}(2,1.5)(0,-0.1)
\put(-0.3,-0.1){$a$}
\put(0.9,-1.4){$b$}
\put(0.9,1.1){$d$}
\put(2,0){\circle*{0.15}}
\put(0,0){\line(1,1){1}}
\put(0,0){\line(1,-1){1}}
\put(2,0){\line(-1,1){1}}
\put(2,0){\line(-1,-1){1}}
\put(1.9,-0.3){$e$}
\put(0.9,-0.1){$u$}
\multiput(1,1)(0.09,0){23}{\circle*{0.02}}
\multiput(1,-1)(0.09,0){23}{\circle*{0.02}}
\end{picture}
}
\put(2,0){
% diagonally oriented face weight 
\begin{picture}(2,1.5)(0,-0.1)
\put(0.9,-1.4){$b$}
\put(0.9,1.1){$d$}
\put(0,0){\line(1,1){1}}
\put(0,0){\line(1,-1){1}}
\put(2,0){\line(-1,1){1}}
\put(2,0){\line(-1,-1){1}}
\put(2.1,-0.1){$c$}
\put(0.7,-0.1){$-u$}
\end{picture}
}
\end{picture}
% adjust lower end
\raisebox{-1.5\unitlength}{}
=
\rho^{qr}(u) \rho^{rq}(-u) \delta_{ac}F^q_{ab} F^r_{ad}.
\end{equation}
For the definition of boundary weights we will need the braid limit of the above bulk 
weights.
The braid limit of the $(p,q)$ weights is defined as
\begin{equation}
\BBpq{p}{q}{a}{b}{c}{d} = 
\lim_{u\to -i \infty}
\frac{1}{\sin^q u} \BWpq{p}{q}{a}{b}{c}{d}{u}.
\end{equation}
The braid limit may be obtained from the above weights by the substitution
\begin{equation}
\lim_{u \to -i \infty} 
\frac{1}{\sin^m u} \gsin{a \pm u}{m} =
\frac{(\pm 1)^m}{\psin{m\l}{m}} e^{\pm i m a \mp i m(m-1) \l/2}.
\end{equation}

We now explain how to define boundary weights, which will realize the different types of
boundary conditions corresponding to the coset description and to the superconformal 
description.
We explain the general method to obtain integrable boundary weights from known boundary 
weights, using fused face weights, as discussed in \cite{BP96}.
We start with a simple initial boundary condition 
\begin{equation}
\DR{q}{0}{b}{c}{a}{u}
=
% diagonally oriented right starting face weight 
\begin{picture}(1.5,1)
\put(0.4,0.1){\line(1,1){0.7}}
\put(0.4,0.1){\line(1,-1){0.7}}
\put(1.1,-0.6){\line(0,1){1.41}}
\put(0.8,0){$u$}
\put(0.1,0){$b$}
\put(1,-1){$c$}
\put(1,0.9){$a$}
\end{picture}
% adjust lower end
\raisebox{-\unitlength}{}
\end{equation}
with weights satisfying the right reflection equation
\begin{equation}
\begin{split}
&\rho^{qr}(u-v+(q-r)\lambda)\rho^{qr}(-u-v-(r-1)\lambda+\mu)\\
&\times \sum_{fg} \BWpq{r}{q}{b}{a}{f}{c}{u-v}
\BWpq{r}{q}{c}{f}{g}{d}{-u-v-(q-1)\lambda+\mu}\\
&\times 
\DR{q}{0}{f}{a}{g}{u}
\DR{r}{0}{d}{g}{e}{u}\\
&=\rho^{rq}(u-v)\rho^{rq}(-u-v-(q-1)\lambda+\mu)\\
&\times \sum_{fg} \BWpq{q}{r}{d}{c}{f}{e}{u-v+(q-r)\lambda}
\BWpq{q}{r}{c}{b}{g}{f}{-u-v-(r-1)\lambda+\mu}\\
&\times 
\DR{q}{0}{f}{g}{e}{u}
\DR{r}{0}{b}{a}{g}{u},
\end{split}
\end{equation}
where $\rho^{qr}(u)$ are model dependent functions.
This is depicted grahically below.
\begin{equation}
\begin{picture}(4.0,2.5)
\put(-2,1){\footnotesize $\rho^{qr}(u-v+(q-r)\lambda)\times$}
\put(-2,1.5){\footnotesize $\rho^{qr}(-u-v-(r-1)\lambda+\mu)\times$}
\put(2,-1){
% right boundary weight 
\begin{picture}(2.5,1)
\put(0.4,0.1){\circle*{0.15}}
\put(0.4,0.1){\line(1,1){1}}
\put(0.4,0.1){\line(1,-1){1}}
\multiput(-0.6,-0.9)(0.09,0){23}{\circle*{0.02}}
\put(1.4,1.1){\circle*{0.15}}
\put(1.4,-0.9){\line(0,1){2}}
\put(1.0,0){\footnotesize $u$}
\put(0.1,0){$b$}
\put(1.3,-1.2){$a$}
\put(1.5,1.0){$g$}
\end{picture}
}
\put(2,1){
% right boundary weight 
\begin{picture}(2.5,1)
\put(0.4,0.1){\line(1,1){1}}
\put(0.4,0.1){\line(1,-1){1}}
\put(1.4,-0.9){\line(0,1){2}}
\put(1.0,0){\footnotesize $v$}
\put(0.2,0.1){$d$}
\put(1.3,1.2){$e$}
\end{picture}
}
\put(0,-1){
% diagonally oriented face weight 
\begin{picture}(2.2,1)
\put(0.4,0.1){\line(1,1){1}}
\put(0.4,0.1){\line(1,-1){1}}
\put(2.4,0.1){\line(-1,1){1}}
\put(2.4,0.1){\line(-1,-1){1}}
\put(1,0.05){\footnotesize $u-v$}
\put(0.1,0){$b$}
\put(1.3,-1.2){$a$}
\put(2.2,-0.3){$f$}
\put(1.2,1.2){$c$}
\end{picture}
}
\put(1,0){
% diagonally oriented face weight 
\begin{picture}(2.2,1)
\put(0.4,0.1){\line(1,1){1}}
\put(0.4,0.1){\line(1,-1){1}}
\put(2.4,0.1){\line(-1,1){1}}
\put(2.4,0.1){\line(-1,-1){1}}
\put(0.8,0){\parbox{1.2cm}{\scriptsize $\mu-u-v\\-(q-1)\lambda$}}
%\put(0.1,0){$a$}
%\put(1.3,-1.3){$b$}
%\put(2.5,0){$c$}
%\put(1.3,1.2){$d$}
\end{picture}
}
\end{picture}
= 
\begin{picture}(4.0,2.5)(-1.5,0)
\put(0.5,-1.5){\footnotesize $\rho^{rq}(u-v)\times$}
\put(-1,-2){\footnotesize $\rho^{rq}(-u-v-(q-1)\lambda+\mu)\times$}
\put(2,-1){
% right boundary weight 
\begin{picture}(2.5,1)
\put(0.4,0.1){\line(1,1){1}}
\put(0.4,0.1){\line(1,-1){1}}
\multiput(-0.6,3.1)(0.09,0){23}{\circle*{0.02}}
\put(1.4,1.1){\circle*{0.15}}
\put(1.4,-0.9){\line(0,1){2}}
\put(1.0,0){\footnotesize $v$}
\put(0.1,-0.1){$b$}
\put(1.3,-1.2){$a$}
\put(1.5,1.0){$g$}
\end{picture}
}
\put(2,1){
% right boundary weight 
\begin{picture}(2.5,1)
\put(0.4,0.1){\line(1,1){1}}
\put(0.4,0.1){\line(1,-1){1}}
\put(1.4,-0.9){\line(0,1){2}}
\put(1.0,0){\footnotesize $u$}
\put(1.3,1.2){$e$}
\end{picture}
}
\put(0,1){
% diagonally oriented face weight 
\begin{picture}(2.2,1)
\put(0.4,0.1){\line(1,1){1}}
\put(0.4,0.1){\line(1,-1){1}}
\put(2.4,0.1){\line(-1,1){1}}
\put(2.4,0.1){\line(-1,-1){1}}
\put(2.4,0.1){\circle*{0.15}}
\put(1,0.05){\footnotesize $u-v$}
\put(0.1,0){$d$}
\put(1.2,-1.1){$c$}
\put(2.3,0.3){$f$}
\put(1.2,1.2){$e$}
\end{picture}
}
\put(1,0){
% diagonally oriented face weight 
\begin{picture}(2.2,1)
\put(0.4,0.1){\line(1,1){1}}
\put(0.4,0.1){\line(1,-1){1}}
\put(2.4,0.1){\line(-1,1){1}}
\put(2.4,0.1){\line(-1,-1){1}}
\put(0.8,0){\parbox{1.2cm}{\scriptsize $\mu-u-v\\-(r-1)\lambda$}}
%\put(0.1,0){$a$}
%\put(1.3,-1.3){$b$}
%\put(2.5,0){$c$}
%\put(1.3,1.2){$d$}
\end{picture}
}
\end{picture}
\raisebox{-2.5cm}{}
\end{equation}

We obtain further integrable boundary weights satisfying the boundary Yang-Baxter
equation by applying $s$-type fusion with the braid bulk weights and $r$-type fusion 
with the face weights \cite{BP01}.
In order to be able to perform the fusion construction, we introduce dangling variables 
for each type of fusion.
This leads to right boundary weights of the form
\begin{equation}
\begin{split}
&\BR{q, (rs)}{b}{c & \gamma_1 & \gamma_0}{a & \alpha_1 & \alpha_0}{u,\xi}
=
\\
&\sum_{\beta_0,\beta_1} \BWpq{r-1}{q}{c}{\gamma_1}{\beta_1}{b}{u-\xi} 
\BWpq{r-1}{q}{b}{\beta_1}{\alpha_1}{a}{\mu-(q-1)\l-u-\xi}\\
&\BBpq{s-1,}{q}{\gamma_1}{\gamma_0}{\beta_0}{\beta_1} 
\BBpq{s-1,}{q}{\beta_1}{\beta_0}{\alpha_0}{\alpha_1}\DR{q}{0}{\beta_0}{\gamma_0}
{\alpha_0}{u},
\end{split}
\end{equation}
graphically depicted by
\begin{equation}
% diagonally oriented right general face weight 
\begin{picture}(2.5,1)
\put(0.4,0.1){\line(1,1){1}}
\put(0.4,0.1){\line(1,-1){1}}
\put(1.4,1.1){\line(1,0){1}}
\put(1.4,-0.9){\line(1,0){1}}
\multiput(2.4,-0.9)(0,0.159){13}{\line(0,1){0.1}}
\multiput(1.9,-0.9)(0,0.159){13}{\line(0,1){0.1}}
\put(0.8,0){$u,\xi$}
\put(0.1,0){$b$}
\put(1.2,-1.2){$c$}
\put(1.7,-1.2){$\gamma_1$}
\put(2.2,-1.2){$\gamma_0$}
\put(1.2,1.2){$a$}
\put(1.7,1.2){$\alpha_1$}
\put(2.2,1.2){$\alpha_0$}
\end{picture}
% adjust lower end
\raisebox{-\unitlength}{}
=
% fusion construction for boundary weights 
\begin{picture}(6,1.5)(-0.5,-0.1)
\put(0,0){\framebox(2,1){\parbox{2cm}{\footnotesize
\centerline{$\mu-(q-1)\lambda$}\centerline{$-u-\xi$}}}}
\put(0,-1){\framebox(2,1){\parbox{2cm}{\footnotesize\centerline{$u-\xi$}}}}
\put(2,0){\framebox(2,1){\parbox{2cm}{}}}
\put(2,-1){\framebox(2,1){\parbox{2cm}{}}}
\put(4,0){\line(1,1){1}}
\put(4,0){\line(1,-1){1}}
\put(5,-1){\line(0,1){2}}
\multiput(4,1)(0.09,0){12}{\circle*{0.02}}
\multiput(4,-1)(0.09,0){12}{\circle*{0.02}}
\put(2,0){\circle*{0.15}}
\put(4,0){\circle*{0.15}}
\put(2.1,-0.3){\small $\beta_1$}
\put(3.6,-0.3){\small $\beta_0$}
\put(-0.3,1.1){$a$}
\put(1.7,1.1){$\alpha_1$}
\put(3.7,1.1){$\alpha_0$}
\put(4.7,1.1){$\alpha_0$}
\put(-0.3,-0.1){$b$}
\put(-0.3,-1.3){$c$}
\put(1.7,-1.3){$\gamma_1$}
\put(3.7,-1.3){$\gamma_0$}
\put(4.7,-1.3){$\gamma_0$}
\put(4.5,-0.1){$u$}
\end{picture}
% adjust lower end
\raisebox{-\unitlength}{}
\end{equation}
Here, $\mu$ and $\xi$ are arbitrary fixed parameters.
The parameter $\mu$ is the crossing parameter (see (\ref{form:crosssym})), which we fix 
to be $\mu=2\lambda$ in our numerical calculations.
The value of the inhomgeneity parameter $\xi$ will later be chosen such that the 
corresponding boundary weights take simple form and are conformally invariant.

The above construction generally introduces dangling variables for the boundary weight.
In some cases, however, the dependence on these variables disappears.
This is, for example, the case for boundary conditions corresponding to the unitary 
minimal models.
Here, the $(r,s)$-type boundary conditions corresponding to the
Virasoro characters of type $(r,s)$ are obtained by starting with a simple ``vacuum'' 
solution with boundary spins $a=c=1$.
Applying fusion $s-1$ times with face weights in the braid limit gives integrable, 
$(1,s)$-type boundary weights.
Since the spin variable of the vacuum weight has only one value, the new boundary weight
is not dependent on the value of this internal spin.
Again, the new $(1,s)$-type weights are diagonal and have only one spin value $a=c=s$.
Repeated $r-1$ times fusion with the full weights leads to the $(r,s)$-type boundary 
conditions.
These are again, by construction, independent of the dangling variable $s$. 

As we will discuss in the next paragraph, above construction yields, for suitable choices
of the starting weights, boundary weights corresponding to each branching function and 
each superconformal character.
In contrast to the boundary weights for the unitary minimal models, these 
boundary weights will however generally depend on internal dangling spins.
The weights satisfy a generalized fused right reflection equation
\begin{equation}\label{form:refeqr}
\begin{split}
&\rho^{qr}(u-v+(q-r)\lambda)\rho^{qr}(-u-v-(r-1)\lambda+\mu)\\
&\times \sum_{fg\gamma_0\gamma_1} \BWpq{r}{q}{b}{a}{f}{c}{u-v}
\BWpq{r}{q}{c}{f}{g}{d}{-u-v-(q-1)\lambda+\mu}\\
&\times \BR{q}{f}{a & \alpha_1 & \alpha_0}{g & \gamma_1 & \gamma_0}{u}
 \BR{r}{d}{g & \gamma_1 & \gamma_0}{e & \epsilon_1 & \epsilon_0}{u}\\
&=\rho^{rq}(u-v)\rho^{rq}(-u-v-(q-1)\lambda+\mu)\\
&\times \sum_{fg\gamma_0\gamma_1} \BWpq{q}{r}{d}{c}{f}{e}{u-v+(q-r)\lambda}
\BWpq{q}{r}{c}{b}{g}{f}{-u-v-(r-1)\lambda+\mu}\\
&\times \BR{q}{f}{g & \gamma_1 & \gamma_0}{e & \epsilon_1 & \epsilon_0}{u}
 \BR{r}{b}{a & \alpha_1 & \alpha_0}{g & \gamma_1 & \gamma_0}{u},
\end{split}
\end{equation}
where $\rho^{qr}(u)$ are model dependent functions.
\begin{equation}
\begin{picture}(5.5,2.5)
\put(-2,1){\footnotesize $\rho^{qr}(u-v+(q-r)\lambda)\times$}
\put(-2,1.5){\footnotesize $\rho^{qr}(-u-v-(r-1)\lambda+\mu)\times$}
\put(2,-1){
% right boundary weight 
\begin{picture}(2.5,1)
\put(0.4,0.1){\circle*{0.15}}
\put(0.4,0.1){\line(1,1){1}}
\put(0.4,0.1){\line(1,-1){1}}
\multiput(-0.6,-0.9)(0.09,0){23}{\circle*{0.02}}
\put(1.4,1.1){\circle*{0.15}}
\put(1.9,1.1){\circle*{0.15}}
\put(2.4,1.1){\circle*{0.15}}
\put(1.4,1.1){\line(1,0){1}}
\put(1.4,-0.9){\line(1,0){1}}
\multiput(2.4,-0.9)(0,0.159){13}{\line(0,1){0.1}}
\multiput(1.9,-0.9)(0,0.159){13}{\line(0,1){0.1}}
\put(1.2,0){\footnotesize $u$}
\put(0.1,0){$b$}
\put(1.2,-1.2){$a$}
\put(1.7,-1.2){$\alpha_1$}
\put(2.2,-1.2){$\alpha_0$}
\put(1.4,0.8){$g$}
\put(1.9,0.8){$\gamma_1$}
\put(2.5,1){$\gamma_0$}
\end{picture}
}
\put(2,1){
% right boundary weight 
\begin{picture}(2.5,1)
\put(0.4,0.1){\line(1,1){1}}
\put(0.4,0.1){\line(1,-1){1}}
\put(1.4,1.1){\line(1,0){1}}
\put(1.4,-0.9){\line(1,0){1}}
\multiput(2.4,-0.9)(0,0.159){13}{\line(0,1){0.1}}
\multiput(1.9,-0.9)(0,0.159){13}{\line(0,1){0.1}}
\put(1.2,0){\footnotesize $v$}
\put(0.2,0.1){$d$}
%\put(1.2,-1.2){$c$}
%\put(1.7,-1.2){$\gamma_1$}
%\put(2.2,-1.2){$\gamma_0$}
\put(1.2,1.2){$e$}
\put(1.7,1.2){$\epsilon_1$}
\put(2.2,1.2){$\epsilon_0$}
\end{picture}
}
\put(0,-1){
% diagonally oriented face weight 
\begin{picture}(2.2,1)
\put(0.4,0.1){\line(1,1){1}}
\put(0.4,0.1){\line(1,-1){1}}
\put(2.4,0.1){\line(-1,1){1}}
\put(2.4,0.1){\line(-1,-1){1}}
\put(1,0.05){\footnotesize $u-v$}
\put(0.1,0){$b$}
\put(1.3,-1.2){$a$}
\put(2.2,-0.3){$f$}
\put(1.2,1.2){$c$}
\end{picture}
}
\put(1,0){
% diagonally oriented face weight 
\begin{picture}(2.2,1)
\put(0.4,0.1){\line(1,1){1}}
\put(0.4,0.1){\line(1,-1){1}}
\put(2.4,0.1){\line(-1,1){1}}
\put(2.4,0.1){\line(-1,-1){1}}
\put(0.8,0){\parbox{1.2cm}{\scriptsize $\mu-u-v\\-(q-1)\lambda$}}
%\put(0.1,0){$a$}
%\put(1.3,-1.3){$b$}
%\put(2.5,0){$c$}
%\put(1.3,1.2){$d$}
\end{picture}
}
\end{picture}
= 
\begin{picture}(5.5,2.5)(-1.5,0)
\put(0.5,-1.5){\footnotesize $\rho^{rq}(u-v)\times$}
\put(-1,-2){\footnotesize $\rho^{rq}(-u-v-(q-1)\lambda+\mu)\times$}
\put(2,-1){
% right boundary weight 
\begin{picture}(2.5,1)
\put(0.4,0.1){\line(1,1){1}}
\put(0.4,0.1){\line(1,-1){1}}
\multiput(-0.6,3.1)(0.09,0){23}{\circle*{0.02}}
\put(1.4,1.1){\circle*{0.15}}
\put(1.9,1.1){\circle*{0.15}}
\put(2.4,1.1){\circle*{0.15}}
\put(1.4,1.1){\line(1,0){1}}
\put(1.4,-0.9){\line(1,0){1}}
\multiput(2.4,-0.9)(0,0.159){13}{\line(0,1){0.1}}
\multiput(1.9,-0.9)(0,0.159){13}{\line(0,1){0.1}}
\put(1.2,0){\footnotesize $v$}
\put(0.1,-0.1){$b$}
\put(1.2,-1.2){$a$}
\put(1.7,-1.2){$\alpha_1$}
\put(2.2,-1.2){$\alpha_0$}
\put(1.4,0.8){$g$}
\put(1.9,0.8){$\gamma_1$}
\put(2.5,1){$\gamma_0$}
\end{picture}
}
\put(2,1){
% right boundary weight 
\begin{picture}(2.5,1)
\put(0.4,0.1){\line(1,1){1}}
\put(0.4,0.1){\line(1,-1){1}}
\put(1.4,1.1){\line(1,0){1}}
\put(1.4,-0.9){\line(1,0){1}}
\multiput(2.4,-0.9)(0,0.159){13}{\line(0,1){0.1}}
\multiput(1.9,-0.9)(0,0.159){13}{\line(0,1){0.1}}
\put(1.2,0){\footnotesize $u$}
%\put(0.2,0.1){$d$}
%\put(1.2,-1.2){$c$}
%\put(1.7,-1.2){$\gamma_1$}
%\put(2.2,-1.2){$\gamma_0$}
\put(1.2,1.2){$e$}
\put(1.7,1.2){$\epsilon_1$}
\put(2.2,1.2){$\epsilon_0$}
\end{picture}
}
\put(0,1){
% diagonally oriented face weight 
\begin{picture}(2.2,1)
\put(0.4,0.1){\line(1,1){1}}
\put(0.4,0.1){\line(1,-1){1}}
\put(2.4,0.1){\line(-1,1){1}}
\put(2.4,0.1){\line(-1,-1){1}}
\put(2.4,0.1){\circle*{0.15}}
\put(1,0.05){\footnotesize $u-v$}
\put(0.1,0){$d$}
\put(1.2,-1.1){$c$}
\put(2.3,0.3){$f$}
\put(1.2,1.2){$e$}
\end{picture}
}
\put(1,0){
% diagonally oriented face weight 
\begin{picture}(2.2,1)
\put(0.4,0.1){\line(1,1){1}}
\put(0.4,0.1){\line(1,-1){1}}
\put(2.4,0.1){\line(-1,1){1}}
\put(2.4,0.1){\line(-1,-1){1}}
\put(0.8,0){\parbox{1.2cm}{\scriptsize $\mu-u-v\\-(r-1)\lambda$}}
%\put(0.1,0){$a$}
%\put(1.3,-1.3){$b$}
%\put(2.5,0){$c$}
%\put(1.3,1.2){$d$}
\end{picture}
}
\end{picture}
\raisebox{-2.5cm}{}
\end{equation}

Similarly, we define left boundary weights, starting from a boundary weight
\begin{equation}
\DL{q}{0}{c}{a}{b}{u}
=
% diagonally oriented right starting face weight 
\begin{picture}(1.5,1)
\put(0.3,-0.6){\line(0,1){1.41}}
\put(1.01,0.1){\line(-1,1){0.7}}
\put(1.01,0.1){\line(-1,-1){0.7}}
\put(0.4,0){$u$}
\put(0.2,-1){$b$}
\put(1.11,0){$c$}
\put(0.2,0.9){$d$}
\end{picture}
% adjust lower end
\raisebox{-\unitlength}{}
\end{equation}
satisfying the left reflection equation
\begin{equation}
\begin{split}
&\rho^{rq}(u-v)\rho^{rq}(-u-v-(q-1)\lambda+\mu)\\
&\times \sum_{fg} \BWpq{q}{r}{f}{a}{b}{c}{u-v+(q-r)\lambda}
\BWpq{q}{r}{g}{f}{c}{d}{-u-v-(r-1)\lambda+\mu}\\
&\times 
\DL{q}{0}{a}{g}{f}{u}
\DL{r}{0}{g}{e}{d}{u}\\
&=\rho^{qr}(u-v+(q-r)\lambda)\rho^{qr}(-u-v-(r-1)\lambda+\mu)\\
&\times \sum_{fg} \BWpq{r}{q}{f}{c}{d}{e}{u-v}
\BWpq{r}{q}{g}{b}{c}{f}{-u-v-(q-1)\lambda+\mu}\\
&\times 
\DL{q}{0}{g}{e}{f}{u}
\DL{r}{0}{a}{g}{b}{u}.
\end{split}
\end{equation}
This is depicted graphically below.
\begin{equation}
\begin{picture}(5,2.5)(1.0,0)
\put(2.5,1.5){\footnotesize $\times \rho^{rq}(u-v)$}
\put(2,2){\footnotesize $\times \rho^{rq}(-u-v-(q-1)\lambda+\mu)$}
\put(0,-1){
% left boundary weight 
\begin{picture}(2.5,1)
\put(2.2,0.1){\line(-1,1){1}}
\put(2.2,0.1){\line(-1,-1){1}}
\put(2.2,0.1){\circle*{0.15}}
\multiput(1.2,-0.9)(0.09,0){23}{\circle*{0.02}}
\put(1.2,-0.9){\line(0,1){2}}
\put(1.4,0){\footnotesize $u$}
\put(1.1,-1.2){$a$}
\end{picture}
}
\put(0,1){
% left boundary weight 
\begin{picture}(2.5,1)
\put(2.2,0.1){\line(-1,1){1}}
\put(2.2,0.1){\line(-1,-1){1}}
\put(1.2,-0.9){\line(0,1){2}}
\put(1.4,0){\footnotesize $v$}
\put(1.2,-0.9){\circle*{0.15}}
\put(0.9,-1.0){$g$}
\put(1.1,1.2){$e$}
\end{picture}
}
\put(1.8,-1){
% diagonally oriented face weight 
\begin{picture}(2.2,1)
\put(0.4,0.1){\line(1,1){1}}
\put(0.4,0.1){\line(1,-1){1}}
\put(2.4,0.1){\line(-1,1){1}}
\put(2.4,0.1){\line(-1,-1){1}}
\put(1,0.05){\footnotesize $u-v$}
%\put(0.1,0){$b$}
\put(1.3,-1.2){$a$}
\put(2.5,0){$b$}
%\put(1.2,1.2){$c$}
\end{picture}
}
\put(0.8,0){
% diagonally oriented face weight 
\begin{picture}(2.2,1)
\put(0.4,0.1){\line(1,1){1}}
\put(0.4,0.1){\line(1,-1){1}}
\put(2.4,0.1){\line(-1,1){1}}
\put(2.4,0.1){\line(-1,-1){1}}
\put(0.8,0){\parbox{1.2cm}{\scriptsize $\mu-u-v\\-(r-1)\lambda$}}
%\put(0.1,0){$a$}
\put(1.2,-1.3){$f$}
\put(2.5,0.1){$c$}
\put(1.3,1.2){$d$}
\end{picture}
}
\end{picture}
= 
\begin{picture}(4.0,2.5)(0,0)
\put(2.5,-1.5){\footnotesize $\times \rho^{qr}(u-v+(q-r)\lambda)$}
\put(2.0,-2){\footnotesize $\times \rho^{qr}(-u-v-(r-1)\lambda+\mu)$}
\put(0,-1){
% left boundary weight 
\begin{picture}(2.5,1)
\put(2.2,0.1){\line(-1,1){1}}
\put(2.2,0.1){\line(-1,-1){1}}
\multiput(1.2,3.1)(0.09,0){23}{\circle*{0.02}}
\put(1.2,-0.9){\line(0,1){2}}
\put(1.4,0){\footnotesize $v$}
\put(1.1,-1.2){$a$}
\end{picture}
}
\put(0,1){
% left boundary weight 
\begin{picture}(2.5,1)
\put(2.2,0.1){\line(-1,1){1}}
\put(2.2,0.1){\line(-1,-1){1}}
\put(2.2,0.1){\circle*{0.15}}
\put(1.2,-0.9){\line(0,1){2}}
\put(1.4,0){\footnotesize $u$}
\put(1.2,-0.9){\circle*{0.15}}
\put(0.9,-1.0){$g$}
\put(1.1,1.2){$e$}
\end{picture}
}
\put(1.8,1){
% diagonally oriented face weight 
\begin{picture}(2.2,1)
\put(0.4,0.1){\line(1,1){1}}
\put(0.4,0.1){\line(1,-1){1}}
\put(2.4,0.1){\line(-1,1){1}}
\put(2.4,0.1){\line(-1,-1){1}}
\put(1,0.05){\footnotesize $u-v$}
%\put(0.1,0){$b$}
%\put(1.3,-1.2){$a$}
\put(2.5,0){$d$}
\put(1.3,1.2){$e$}
\end{picture}
}
\put(0.8,0){
% diagonally oriented face weight 
\begin{picture}(2.2,1)
\put(0.4,0.1){\line(1,1){1}}
\put(0.4,0.1){\line(1,-1){1}}
\put(2.4,0.1){\line(-1,1){1}}
\put(2.4,0.1){\line(-1,-1){1}}
\put(0.8,0){\parbox{1.2cm}{\scriptsize $\mu-u-v\\-(q-1)\lambda$}}
%\put(0.1,0){$a$}
\put(1.3,-1.3){$b$}
\put(2.5,0){$c$}
\put(1.3,1.4){$f$}
\end{picture}
}
\end{picture}
\raisebox{-2.5cm}{}
\end{equation}

We obtain integrable boundary weights by subsequently applying $s$-type fusion 
and $r$-type fusion.
The new left boundary weights are given explicitly by
\begin{equation}
\begin{split}
&\BL{q, (rs)}{ \gamma_0 & \gamma_1 &c}{ \alpha_0 & \alpha_1 &a}{b}{u,\xi} = \\
&\sum_{\beta_0,\beta_1}\DL{q}{0}{\gamma_0}{\alpha_0}{\beta_0}{u}
\BBpq{s-1,}{q}{\gamma_0}{\gamma_1}{\beta_1}{\beta_0} 
\BBpq{s-1,}{q}{\beta_0}{\beta_1}{\alpha_1}{\alpha_0} \\
& \BWpq{r-1}{q}{\gamma_1}{c}{b}{\beta_1}{\mu-(q-1)\l-u-\xi} 
\BWpq{r-1}{q}{\beta_1}{b}{a}{\alpha_1}{u-\xi},
\end{split}
\end{equation}
depicted graphically as
\begin{equation}
% diagonally oriented left general face weight 
\begin{picture}(2.5,1)
\put(2.2,0.1){\line(-1,1){1}}
\put(2.2,0.1){\line(-1,-1){1}}
\put(0.2,1.1){\line(1,0){1}}
\put(0.2,-0.9){\line(1,0){1}}
\multiput(0.7,-0.9)(0,0.159){13}{\line(0,1){0.1}}
\multiput(0.2,-0.9)(0,0.159){13}{\line(0,1){0.1}}
\put(1.2,0){$u,\xi$}
\put(2.3,0){$b$}
\put(1.0,-1.2){$c$}
\put(0,-1.2){$\gamma_0$}
\put(0.5,-1.2){$\gamma_1$}
\put(1,1.2){$a$}
\put(0,1.2){$\alpha_0$}
\put(0.5,1.2){$\alpha_1$}
\end{picture}
% adjust lower end
\raisebox{-\unitlength}{}
=
% fusion construction for boundary weights 
\begin{picture}(6,1.5)(0,-0.1)
\put(1,0){\framebox(2,1){\parbox{2cm}{}}}
\put(1,-1){\framebox(2,1){\parbox{2cm}{}}}
\put(3,-1){\framebox(2,1){\parbox{2cm}{\footnotesize
\centerline{$\mu-(q-1)\lambda$}\centerline{$-u-\xi$}}}}
\put(3,0){\framebox(2,1){\parbox{2cm}{\footnotesize\centerline{$u-\xi$}}}}
\put(1,0){\line(-1,1){1}}
\put(1,0){\line(-1,-1){1}}
\put(0,-1){\line(0,1){2}}
\multiput(0,1)(0.09,0){12}{\circle*{0.02}}
\multiput(0,-1)(0.09,0){12}{\circle*{0.02}}
\put(3,0){\circle*{0.15}}
\put(1,0){\circle*{0.15}}
\put(1.1,-0.3){\small $\beta_0$}
\put(2.6,-0.3){\small $\beta_1$}
\put(5.1,1.1){$a$}
\put(2.7,1.1){$\alpha_1$}
\put(0.8,1.1){$\alpha_0$}
\put(-0.3,1.1){$\alpha_0$}
\put(5.1,-0.1){$b$}
\put(5.1,-1.3){$c$}
\put(2.7,-1.3){$\gamma_1$}
\put(0.8,-1.3){$\gamma_0$}
\put(-0.3,-1.3){$\gamma_0$}
\put(0.3,-0.1){$u$}
\end{picture}
% adjust lower end
\raisebox{-1.5\unitlength}{}
\end{equation}
These boundary weights satisfy a generalized fused left reflection equation
\begin{equation}\label{form:refeql}
\begin{split}
&\rho^{rq}(u-v)\rho^{rq}(-u-v-(q-1)\lambda+\mu)\\
&\times \sum_{fg\gamma_0\gamma_1} \BWpq{q}{r}{f}{a}{b}{c}{u-v+(q-r)\lambda}
\BWpq{q}{r}{g}{f}{c}{d}{-u-v-(r-1)\lambda+\mu}\\
&\times \BL{q}{\alpha_0 & \alpha_1 &a}{\gamma_0 & \gamma_1 &g}{f}{u}
 \BL{r}{\gamma_0 & \gamma_1 & g}{\epsilon_0 & \epsilon_1 & e}{d}{u}\\
&=\rho^{qr}(u-v+(q-r)\lambda)\rho^{qr}(-u-v-(r-1)\lambda+\mu)\\
&\times \sum_{fg\gamma_0\gamma_1} \BWpq{r}{q}{f}{c}{d}{e}{u-v}
\BWpq{r}{q}{g}{b}{c}{f}{-u-v-(q-1)\lambda+\mu}\\
&\times \BL{q}{\gamma_1 & \gamma_0 &g}{\epsilon_1 & \epsilon_0 & e}{f}{u}
 \BL{r}{\alpha_1 & \alpha_0 & a}{\gamma_1 & \gamma_0 & g}{b}{u}.
\end{split}
\end{equation}
\begin{equation}
\begin{picture}(6,2.5)
\put(2.5,1.5){\footnotesize $\times \rho^{rq}(u-v)$}
\put(2,2){\footnotesize $\times \rho^{rq}(-u-v-(q-1)\lambda+\mu)$}
\put(0,-1){
% left boundary weight 
\begin{picture}(2.5,1)
\put(2.2,0.1){\line(-1,1){1}}
\put(2.2,0.1){\line(-1,-1){1}}
\put(2.2,0.1){\circle*{0.15}}
\put(0.2,1.1){\line(1,0){1}}
\put(0.2,-0.9){\line(1,0){1}}
\multiput(1.2,-0.9)(0.09,0){23}{\circle*{0.02}}
\multiput(0.7,-0.9)(0,0.159){13}{\line(0,1){0.1}}
\multiput(0.2,-0.9)(0,0.159){13}{\line(0,1){0.1}}
\put(1.2,0){\footnotesize $u$}
%\put(2.3,0){$b$}
\put(1.0,-1.2){$a$}
\put(0,-1.2){$\alpha_0$}
\put(0.5,-1.2){$\alpha_1$}
%\put(1,1.2){$a$}
%\put(0,1.2){$\alpha_0$}
%\put(0.5,1.2){$\alpha_1$}
\end{picture}
}
\put(0,1){
% left boundary weight 
\begin{picture}(2.5,1)
\put(2.2,0.1){\line(-1,1){1}}
\put(2.2,0.1){\line(-1,-1){1}}
\put(0.2,1.1){\line(1,0){1}}
\put(0.2,-0.9){\line(1,0){1}}
\multiput(0.7,-0.9)(0,0.159){13}{\line(0,1){0.1}}
\multiput(0.2,-0.9)(0,0.159){13}{\line(0,1){0.1}}
\put(1.2,0){\footnotesize $v$}
%\put(2.3,0){$b$}
\put(1.2,-0.9){\circle*{0.15}}
\put(0.7,-0.9){\circle*{0.15}}
\put(0.2,-0.9){\circle*{0.15}}
\put(1,-1.2){$g$}
\put(-0.3,-1.0){$\gamma_0$}
\put(0.3,-1.2){$\gamma_1$}
\put(1,1.2){$e$}
\put(0,1.2){$\epsilon_0$}
\put(0.5,1.2){$\epsilon_1$}
\end{picture}
}
\put(1.8,-1){
% diagonally oriented face weight 
\begin{picture}(2.2,1)
\put(0.4,0.1){\line(1,1){1}}
\put(0.4,0.1){\line(1,-1){1}}
\put(2.4,0.1){\line(-1,1){1}}
\put(2.4,0.1){\line(-1,-1){1}}
\put(1,0.05){\footnotesize $u-v$}
%\put(0.1,0){$b$}
\put(1.3,-1.2){$a$}
\put(2.5,0){$b$}
%\put(1.2,1.2){$c$}
\end{picture}
}
\put(0.8,0){
% diagonally oriented face weight 
\begin{picture}(2.2,1)
\put(0.4,0.1){\line(1,1){1}}
\put(0.4,0.1){\line(1,-1){1}}
\put(2.4,0.1){\line(-1,1){1}}
\put(2.4,0.1){\line(-1,-1){1}}
\put(0.8,0){\parbox{1.2cm}{\scriptsize $\mu-u-v\\-(r-1)\lambda$}}
%\put(0.1,0){$a$}
\put(1.2,-1.3){$f$}
\put(2.5,0.1){$c$}
\put(1.3,1.2){$d$}
\end{picture}
}
\end{picture}
= 
\begin{picture}(5.5,2.5)(-0.5,0)
\put(2.5,-1.5){\footnotesize $\times \rho^{qr}(u-v+(q-r)\lambda)$}
\put(2.0,-2){\footnotesize $\times \rho^{qr}(-u-v-(r-1)\lambda+\mu)$}
\put(0,-1){
% left boundary weight 
\begin{picture}(2.5,1)
\put(2.2,0.1){\line(-1,1){1}}
\put(2.2,0.1){\line(-1,-1){1}}
\put(0.2,1.1){\line(1,0){1}}
\put(0.2,-0.9){\line(1,0){1}}
\multiput(1.2,3.1)(0.09,0){23}{\circle*{0.02}}
\multiput(0.7,-0.9)(0,0.159){13}{\line(0,1){0.1}}
\multiput(0.2,-0.9)(0,0.159){13}{\line(0,1){0.1}}
\put(1.2,0){\footnotesize $v$}
%\put(2.3,0){$b$}
\put(1.0,-1.2){$a$}
\put(0,-1.2){$\alpha_0$}
\put(0.5,-1.2){$\alpha_1$}
%\put(1,1.2){$a$}
%\put(0,1.2){$\alpha_0$}
%\put(0.5,1.2){$\alpha_1$}
\end{picture}
}
\put(0,1){
% left boundary weight 
\begin{picture}(2.5,1)
\put(2.2,0.1){\line(-1,1){1}}
\put(2.2,0.1){\line(-1,-1){1}}
\put(0.2,1.1){\line(1,0){1}}
\put(0.2,-0.9){\line(1,0){1}}
\put(2.2,0.1){\circle*{0.15}}
\multiput(0.7,-0.9)(0,0.159){13}{\line(0,1){0.1}}
\multiput(0.2,-0.9)(0,0.159){13}{\line(0,1){0.1}}
\put(1.2,0){\footnotesize $u$}
%\put(2.3,0){$b$}
\put(1.2,-0.9){\circle*{0.15}}
\put(0.7,-0.9){\circle*{0.15}}
\put(0.2,-0.9){\circle*{0.15}}
\put(1,-1.2){$g$}
\put(-0.3,-1.0){$\gamma_0$}
\put(0.3,-1.2){$\gamma_1$}
\put(1,1.2){$e$}
\put(0,1.2){$\epsilon_0$}
\put(0.5,1.2){$\epsilon_1$}
\end{picture}
}
\put(1.8,1){
% diagonally oriented face weight 
\begin{picture}(2.2,1)
\put(0.4,0.1){\line(1,1){1}}
\put(0.4,0.1){\line(1,-1){1}}
\put(2.4,0.1){\line(-1,1){1}}
\put(2.4,0.1){\line(-1,-1){1}}
\put(1,0.05){\footnotesize $u-v$}
%\put(0.1,0){$b$}
%\put(1.3,-1.2){$a$}
\put(2.5,0){$d$}
\put(1.3,1.2){$e$}
\end{picture}
}
\put(0.8,0){
% diagonally oriented face weight 
\begin{picture}(2.2,1)
\put(0.4,0.1){\line(1,1){1}}
\put(0.4,0.1){\line(1,-1){1}}
\put(2.4,0.1){\line(-1,1){1}}
\put(2.4,0.1){\line(-1,-1){1}}
\put(0.8,0){\parbox{1.2cm}{\scriptsize $\mu-u-v\\-(q-1)\lambda$}}
%\put(0.1,0){$a$}
\put(1.3,-1.3){$b$}
\put(2.5,0){$c$}
\put(1.3,1.4){$f$}
\end{picture}
}
\end{picture}
\raisebox{-2.5cm}{}
\end{equation}
The fused double-row transfer matrices are defined by
\begin{equation}
\begin{split}
&< \alpha_L^0,\alpha_L^1,a_1,\ldots,a_{N+1},\alpha_R^1,\alpha_R^0| {\bf
D}^{pq}(u,\xi_1,\xi_2) | 
\beta_L^0,\beta_L^1, b_1,\ldots,b_{N+1}, \beta_R^1, \beta_R^0 >\\
& = \sum_{c_1\ldots c_{N+1}} 
\BL{q}{\alpha_L^0 & \alpha_L^1 & a_1}{\beta_L^0 & \beta_L^1 &
b_1}{c_1}{-u-(q-1)\lambda+\mu,\xi_1} \\
&\times \left[ \prod_{j=1}^N  
\BWpq{p}{q}{a_j}{a_{j+1}}{c_{j+1}}{c_j}{u}
\BWpq{p}{q}{c_j}{c_{j+1}}{b_{j+1}}{b_j}{-u-(q-1)\lambda+\mu}\right]\\
& \times \BR{q}{c_{N+1}}{a_{N+1} & \alpha_R^1 & \alpha_R^0}
{b_{N+1} & \beta_R^1 & \beta_R^0}{u,\xi_2}\\
& =
% diagonally oriented left general face weight 
\begin{picture}(10.5,1.5)
\put(2.2,0.1){\line(-1,1){1}}
\put(2.2,0.1){\line(-1,-1){1}}
\put(0.2,1.1){\line(1,0){1}}
\put(0.2,-0.9){\line(1,0){1}}
\multiput(0.7,-0.9)(0,0.159){13}{\line(0,1){0.1}}
\multiput(0.2,-0.9)(0,0.159){13}{\line(0,1){0.1}}
\put(0.8,0){\scriptsize \parbox{1.2cm}{$\mu-u-$\\$(q-1)\lambda,$\\\centerline{$\xi$}}}
\put(2.3,-0.2){\footnotesize $c_1$}
\put(1.0,-1.2){\footnotesize $a_1$}
\put(0,-1.2){\footnotesize $\alpha_L^0$}
\put(0.5,-1.2){\footnotesize $\alpha_L^1$}
\put(1,1.2){\footnotesize $b_1$}
\put(0,1.2){\footnotesize $\beta_L^0$}
\put(0.5,1.2){\footnotesize $\beta_L^1$}
\multiput(1.2,1.1)(0.09,0){12}{\circle*{0.02}}
\multiput(1.2,-0.9)(0.09,0){12}{\circle*{0.02}}
\put(2,1.2){\footnotesize $b_1$}
\put(3,1.2){\footnotesize $b_2$}
\put(4,1.2){\footnotesize $b_3$}
\put(6,1.2){\footnotesize $b_N$}
\put(6.8,1.2){\footnotesize $b_{N+1}$}
\put(3.3,-0.2){\footnotesize $c_2$}
\put(4.3,-0.2){\footnotesize $c_3$}
\put(5.7,-0.2){\footnotesize $c_N$}
\put(2,-1.2){\footnotesize $a_1$}
\put(3,-1.2){\footnotesize $a_2$}
\put(4,-1.2){\footnotesize $a_3$}
\put(6,-1.2){\footnotesize $a_N$}
\put(6.8,-1.2){\footnotesize $a_{N+1}$}
\put(2.2,0.1){\circle*{0.15}}
\put(3.2,0.1){\circle*{0.15}}
\put(4.2,0.1){\circle*{0.15}}
\put(6.2,0.1){\circle*{0.15}}
\put(7.2,0.1){\circle*{0.15}}
\put(2.6,-0.55){\footnotesize $u$}
\put(3.6,-0.55){\footnotesize $u$}
\put(6.6,-0.55){\footnotesize $u$}
\put(2.2,0.1){\framebox(1,1){\parbox{1cm}{\tiny
\centerline{$\mu-u-$}\centerline{$(q-1)\lambda$}}}}
\put(2.2,-0.9){\framebox(1,1){\parbox{1cm}{}}}
\put(3.2,0.1){\framebox(1,1){\parbox{1cm}{\tiny
\centerline{$\mu-u-$}\centerline{$(q-1)\lambda$}}}}
\put(3.2,-0.9){\framebox(1,1){\parbox{1cm}{}}}
\put(4.2,0.1){\framebox(2,1){\parbox{1cm}{}}}
\put(4.2,-0.9){\framebox(2,1){\parbox{1cm}{}}}
\put(6.2,0.1){\framebox(1,1){\parbox{1cm}{\tiny
\centerline{$\mu-u-$}\centerline{$(q-1)\lambda$}}}}
\put(6.2,-0.9){\framebox(1,1){\parbox{1cm}{}}}
\multiput(7.2,1.1)(0.09,0){12}{\circle*{0.02}}
\multiput(7.2,-0.9)(0.09,0){12}{\circle*{0.02}}
% diagonally oriented right general face weight 
\put(7.2,0.1){\line(1,1){1}}
\put(7.2,0.1){\line(1,-1){1}}
\put(8.2,1.1){\line(1,0){1}}
\put(8.2,-0.9){\line(1,0){1}}
\multiput(9.2,-0.9)(0,0.159){13}{\line(0,1){0.1}}
\multiput(8.7,-0.9)(0,0.159){13}{\line(0,1){0.1}}
\put(7.6,0){\footnotesize $u,\xi$}
\put(6.45,-0.2){\footnotesize $c_{N+1}$}
\put(7.7,-1.2){\footnotesize $a_{N+1}$}
\put(8.5,-1.2){\footnotesize $\alpha_R^1$}
\put(9.0,-1.2){\footnotesize $\alpha_R^0$}
\put(7.7,1.2){\footnotesize $b_{N+1}$}
\put(8.5,1.2){\footnotesize $\beta_R^1$}
\put(9.0,1.2){\footnotesize $\beta_R^0$}
\end{picture}
% adjust lower end
\raisebox{-1.5\unitlength}{}
\end{split}
\end{equation}

The fused double-row transfer matrices form a commuting family
\begin{equation}
{\bf D}^{pq}(u){\bf D}^{pq}(v)={\bf D}^{pq}(v){\bf D}^{pq}(u).
\end{equation}
This can be shown by using the fused Yang-Baxter equation (\ref{form:YBE}), 
inversion relation (\ref{form:inv}), and the generalized reflection equations
(\ref{form:refeqr}) and (\ref{form:refeql}) 
in the diagram proof given in \cite{BPO95}.
It can also be shown by similar arguments involving boundary crossing equations,
which we do not give here, that the fused double-row transfer matrices satisfy 
crossing symmetry
\begin{equation}\label{form:crosssym}
{\bf D}^{pq}(u)={\bf D}^{pq}(-u-(q-1)\lambda+\mu).
\end{equation}

\subsection{Finite-size corrections}

The properties of the lattice models connect to the data of the associated
conformal field theories through the finite-size corrections to the eigenvalues of the
double-row transfer matrices.
Let us denote the double-row transfer matrix with boundary coset or supersymmetric
labels $i$ on the left and $j$ on the right by ${\bf D}_{i|j}$.
If we write the eigenvalues of ${\bf D}_{i|j}$ as
\begin{equation}
D_n(u)=\exp(-E_n(u)), \quad n=0,1,2,\ldots
\end{equation}
then the finite size corrections take the form
\begin{equation}\label{form:levels}
E_n(u) = 2 N f(u) +f_{i|j}(u)
 + \frac{2\pi \sin \theta}{N}
\left( -\frac{c}{24} + \Delta_n+k_n\right) + o\left( \frac{1}{N} \right), 
\qquad k_n \in \mathbb{N},
\end{equation}
where $f(u)$ is the bulk free energy, $f_{i|j}(u)$ is the boundary free energy, 
$c$ is the central charge, $\Delta_n$ is a conformal weight and
the anisotropy angle is given by
\begin{equation}
\theta = g u,
\end{equation}
where $g$ is the Coxter number of the graph $A_{g-1}$.

The bulk and boundary free energies can be computed using inversion relations
\cite{BPO95, MP01}.
This we do not do since we are interested only in the conformal partition functions.
Removing the bulk and boundary contributions to the partition function on a cylinder 
leads to the conformal partition function $Z_{i|j}(q)$ with left and right boundaries
$i$ and $j$.
For the superconformal theories, this can be expressed as a linear form in
superconformal characters
\begin{equation}\label{form:frules}
Z_{i|j} = 
\sum_k  n_{i,j}^{k} \chi_k(q).
\end{equation}
where the fusion coefficients $n_{ij}^{\mbox{ } k}\in \mathbb{Z}$ 
give the operator content, and $k$ has to be summed over an appropriate domain  
(\ref{form:supercyl}).
For the coset boundary weights to be defined below, the cylinder partition functions 
are of the form
\begin{equation}\label{form:cosetcyl}
Z_{(r_1,s_1,l_1)|(r_2,s_2,l_2)} = 
\sum_{r=1}^{m-1} \sum_{s=1}^{m+1} \sum_{l=0}^2 
n_{(r_1,s_1,l_1),(r_2,s_2,l_2)}^{(r,s,l)} b_{rs}^{(l)}(q),
\end{equation}
where the coefficients are given by the Verlinde formula (\ref{form:cosetVer}).
It can be checked that this coincides with (\ref{form:coset2cyl}), if the the boundary
condition corresponding to the fixed point branching function is expressed as a linear 
combination of the two fixed point branching fields according to (\ref{form:brfpfield}).

With the introduction of two dangling variables per boundary weight, we are effectively
dealing with four boundary conditions, such that each eigenvector is fourfold 
degenerate. 
For $M$ double rows the modular parameter is
\begin{equation}
q=\exp(2\pi i \tau), \qquad \tau= i \frac{M}{N}\sin \theta,
\end{equation}
where $M/N$ is the aspect ratio of the cylinder.

\section{Coset and superconformal boundary weights}

In this section, we define the integrable coset boundary weights and 
integrable superconformal boundary weights.
Since it is not obvious from the construction of the weights how to identify the
$(r,s)$ labels of the fusion construction with the $(r,s)$ labels in the Kac tables,
we have to make this identification from numerical data.
In the sequel we focus on right boundary weights.
Since the left boundary weights are defined in the same manner,
we do not give the corresponding expressions here.

We first give the boundary weights corresponding to the branching functions  
$b_{r,s}^{(l)}(q)$, which we will denote by $\BR{(r,s,l)}{b}{c & \gamma_1 &\gamma_0}
{a & \alpha_1 & \alpha_0}{u}$.
The starting weight is given by
\begin{equation}
\DR{2}{0}{b}{c}{a}{u} = \delta_{a,1}\delta_{c,1}\delta_{b,3}.
\end{equation}
This weight gives the vacuum character of the above models and  
generalizes the vacuum boundary condition of the unfused $A$ models \cite{BP01}
to fusion level 2.
It is the coset vacuum.
We use this boundary weight on the left of the double-row transfer matrix.
Since the cylinder partition function reduces to a single branching function, it is easy
to identify labels of boundary weights on the right with their corresponding Kac labels.
It can be checked that the $(r,1)$ weights obtained from the above starting weight 
correspond to the weights (6.32) in \cite{BPO95}.

As it turns out, the different boundary weights for different sectors $l$ correspond to 
different choices of the inhomogeneity parameter $\xi$.
For the crossing parameter fixed to $\mu=2\lambda$, we have explicitly
\begin{eqnarray}
\BR{(r,s,0)}{b}{c & \gamma_1 &\gamma_0}{a & \alpha_1 & \alpha_0}{u} & = & 
\BR{2,(r,s)}{b}{c & \gamma_1 &\gamma_0}{a & \alpha_1 & \alpha_0}{u,\xi=5\lambda/2}\\
\BR{(r,s,1)}{b}{c & \gamma_1 &\gamma_0}{a & \alpha_1 & \alpha_0}{u} & = & 
\BR{2,(r+1,s)}{b}{c & \gamma_1 &\gamma_0}{a & \alpha_1 & \alpha_0}{u,\xi=3\lambda/2} 
\nonumber\\
\BR{(r,s,2)}{b}{c & \gamma_1 &\gamma_0}{a & \alpha_1 & \alpha_0}{u} & = & 
\BR{2,(r+2,s)}{b}{c & \gamma_1 &\gamma_0}{a & \alpha_1 & \alpha_0}{u,\xi=-\lambda} 
\nonumber
\end{eqnarray}
As explained before, the above weights are do not in fact depend on the dangling 
variables.
We emphasize that this construction yields coset boundary weights for the lattice model
with odd spin values.
This is no restriction for $m$ odd, since the lattice model on the other subgraph is
identical.
For $m$ even, we observed that the above boundary weights on the {\it even} subgraph
represent {\it superconformal} boundary conditions!
Since this phenomenon does not occur for $m$ odd, we have to introduce the more general 
construction given above in order to obtain superconformal boundary weights in this case 
as well.

The boundary weights realizing superconformal boundary conditions are denoted by
\begin{equation}
\BR{X (r,s)}{b}{c & \gamma_1 &\gamma_0}{a & \alpha_1 & \alpha_0}{u},
\end{equation}
where $X \in \{NS, \widetilde{NS}, R\}$ stands for the 
Neveu-Schwarz sector, Neveu-Schwarz tilda sector or for the Ramond sector, respectively. 
The starting weight in the Neveu-Schwarz sector is given by
\begin{equation}
\DR{NS}{0}{b}{c}{a}{u}= \left[ \BR{(1,1,0)}{b}{c & 1 & 1}{a & 1 & 1}{u,-\frac{\l}{2}} + 
\BR{(3,1,0)}{b}{a & 1 & 1}{c & 1 & 1}{u,-\frac{\l}{2}} \right]/2.
\end{equation}
This weight is the superconformal vacuum.
It satisfies the right reflection equation.
This is due to the fact that each summand satisfies the reflection equation by
construction, and they are both diagonal with different nonzero spin values.
Therefore, the sum in the reflection equation decouples into the two separate reflection
equations.
General $(r,s)$-type boundary weights are obtained by $(r,s)$-fusion with 
inhomogeneity $\xi=5\l/2$.
At the isotropic point $u=\l/2$, the above weight simplifies to
\begin{equation}
\DR{NS}{0}{b}{c}{a}{\frac{\l}{2}} = 
h(\delta_{a,1}\delta_{c,1}\delta_{b,3}+\delta_{a,3}\delta_{c,3}\delta_{b,1}),
\end{equation}
where $h$ is a constant.
The starting weight in the Neveu-Schwarz tilda sector is given by
\begin{equation}
\DR{\widetilde{NS}}{0}{b}{c}{a}{u}= \left[ \BR{(1,1,0)}{b}{c & 1 & 1}{a & 1 & 1}{u,-\frac{\l}{2}} - 
\BR{(3,1,0)}{b}{c & 1 & 1}{a & 1 & 1}{u,-\frac{\l}{2}} \right]/2.
\end{equation}
General $(r,s)$-type boundary weights are obtained by $(r,s)$-fusion with 
inhomogeneity $\xi=5\l/2$.
The superconformal boundary weights in the Ramond sector are given by
\begin{equation}
\DR{R}{0}{b}{c}{a}{u}= \left[ \BR{(2,1,0)}{b}{c & 1 & 1}{a & 1 & 1}{u,-\frac{\l}{2}} +
\BR{(4,1,0)}{b}{c & 1 & 1}{a & 1 & 1}{u,-\frac{\l}{2}} \right]/2.
\end{equation}
General $(r,s)$-type boundary weights are obtained by $(r,s)$-fusion with 
inhomogeneity $\xi=3\l/2$.
These choices of the superconformal boundary weights correspond to the
relation between the branching functions and superconformal characters 
(\ref{form:braco}). 
The labels $(r,s)$ which appear are the superconformal labels in the Kac Table.

\section{Numerical spectra}

Here, we describe our numerical procedure which led to the identification of boundary
conditions presented in the previous chapter.
We have tested our predictions for the models $A_4$ and $A_5$, separately for the coset 
boundary weights and superconformal boundary weights.

We first remind that the conformal part of the finite size corrections to the 
{\em periodic} transfer matrices leads to superconformal torus partition functions.
We confirm for $A_4$ and $A_5$ the form (\ref{form:torcoset}) with $a=1/2$ and $b=0$, 
i.e. the partition sum is just the sum of the square all branching functions.
For $A_4$, and generally for $A_L$ with $L$ even, this is the sum of two identical 
theories.
For $A_5$, we observe that the conformal partition sum on the odd subgraph is given by 
(\ref{form:torcoset}) with $a=1/4$ and $b=1/2$, whereas is is given on the even
sublattice by $a=1/4$ and $b=-1/2$.
This indicates the presence of the trivial fixed point field (corresponding to $b
\neq 0$) for each subgraph.
As mentioned above, our construction however does not yield a lattice
realization of this boundary field.

For the coset boundary weights, which do not depend on dangling variables, we were able
to compute double row transfer matrices up to 16 faces for $A_4$ and up to 11 faces for
$A_5$.
Due to the introduction of dangling variables, double-row transfer matrices of
superconformal boundary weights generally can only be computed for much smaller lattice
sizes, typically up to 5 faces for $A_5$.
For $(r,1)$ or $(1,s)$ type superconformal boundary weights, however, the situation 
can be improved, since the dependence on one dangling variable is trivial and 
may be disregarded.

The $A_4$ model, which has central charge $c=7/10$, can be related to the 
tricritical hard square and tricritical Ising model.
It can be alternatively realized as a unitary minimal model from the (unfused) 
\ade lattice model $A_4$.
The corresponding conformal boundary conditions have been given previously in 
\cite{BP01}.
The coset boundary conditions agree with the conformal boundary conditions.
This is related to the fact that, for this model, the branching functions are
just the Virasoro characters of the model ${\cal M}(7/10)$.

The predictions from conformal field theory manifest themselves in the level spacings
and degeneracies of the double-row transfer matrix eigenvalues in the large $N$ limit,
cf. (\ref{form:levels}).
We have chosen $u=\lambda/2$ such that the sine factor reduces to unity.
For the double-row transfer matrix at fusion level $(2,2)$, which is the case of interest
for our numerics, we achieved this by choosing the isotropic point $u_c=(\mu-\lambda)/2$,
in which case ${\bf D}^{22}(u)={\bf D}^{22}(\mu-\lambda-u)$, and setting 
$\mu=2\lambda$.

First, we have computed double-row transfer matrices with the vacuum weight on the left
and a general boundary weight on the right.
In this case, the cylinder partition function reduces to a single character, according
to the fusion rules (\ref{form:frules}).
In order to check for conformal dimension from given transfer matrix data, we computed
reduced energies by subtracting the contributions from the bulk free energy, from the
boundary free energy and from the central charge according to (\ref{form:levels}). 
We then plotted the largest reduced eigenvalue of the transfer matrix against $1/N$ and 
extrapolated the sequence of numbers to $N=\infty$.
In all cases, we obtained agreement with the theoretical value of $\Delta$ within 
numerical accuracy.

The same method has been applied in order to test the exponents and degeneracies of the 
eigenvalues of the double-row transfer matrix, which are given by the expansion of the
characters in powers of $q$ in the large $N$ limit.
As example, we extract the superconformal vacuum character for $A_4$.
It has a series expansion
\begin{equation}
\chi^{NS}_{1,1}(q)=q^{-7/240}(1+q^{\frac{3}{2}}+q^2+q^{\frac{5}{2}}+q^3+2q^{\frac{7}{2}}+
2q^4+2q^{\frac{9}{2}}+{\cal O}(q^5))
\end{equation}
We have computed the double-row transfer matrix with superconformal vacuum weights on 
the left and on the right up to 15 faces. 
(Note that the dependence of the boundary weight on dangling variables is trivial.)
A polynomial extrapolation of the first ten reduced eigenvalues from lattice sizes 10 to 
15 to $N=\infty$ yields the exponents shown in the table.
\\ 
\\
\begin{tabular}{|r|cccccccccc|}
\hline
energy  & 1 & 2 & 3 & 4 & 5 & 6 & 7 & 8 & 9 & 10 \\
\hline
data  & {\small 0.000} & {\small 1.495} & {\small 1.994} & {\small 2.492} & 
{\small 2.993} & {\small 3.489} & {\small 3.504} & {\small 3.990} & {\small 4.077} & 
{\small 4.501} \\
theory & $\raisebox{-1.5ex}{}$ 0 & $\frac{3}{2}$ & 2 & $\frac{5}{2}$ & 3 &
$\frac{7}{2}$ & $\frac{7}{2}$ & 4 & 4 & $\frac{9}{2}$\\
\hline
\end{tabular}
\\ 
\\
In order to test the predictions for fusion rules, we put different boundary weights to
the right and to the left and tested for the correct cylinder partition function by
examining the first ten eigenvectors of the double-row transfer matrix.
In each case tested, we find agreement between theory and prediction within numerical
accuracy.
We discuss a typical example of the coset theory:
The cylinder partition function of the $A_5$ model with left and right boundary weights
of type $(2,3,1)$ is given by (\ref{form:cosetcyl})
\begin{eqnarray}
Z_{(2,3,1)|(2,3,1)} &=& 2 \left(
b^{(0)}_{1,1} +  b^{(0)}_{1,3} +  b^{(0)}_{1,5} +  b^{(0)}_{3,1}+
b^{(0)}_{3,3} +  b^{(0)}_{3,5} \right) \\
&=& q^{-1/24}(2+2q^{\frac{1}{6}}+2q^{\frac{2}{3}}+2q+2q^{\frac{7}{6}}+4q^{\frac{3}{2}}
+{\cal O}(q^{\frac{5}{3}})).
\end{eqnarray}
We have computed the double-row transfer matrix of this model up to 9 faces. 
A polynomial extrapolation of the first ten reduced eigenvalues from lattice sizes 4 to 
9 to $N=\infty$ yields the exponents shown in the table.
\\ 
\\
\begin{tabular}{|r|cccccccccc|}
\hline
energy  & 1 & 2 & 3 & 4 & 5 & 6 & 7 & 8 & 9 & 10 \\
\hline
data  & {\small 0.000} & {\small 0.000} & {\small 0.168} & {\small 0.168} & 
{\small 0.675} & {\small 0.675} & {\small 1.012} & {\small 1.012} & {\small 1.183} & 
{\small 1.183} \\
theory & $\raisebox{-1.5ex}{}$ 0 & 0 & $\frac{1}{6}$ & $\frac{1}{6}$ & $\frac{2}{3}$ &
$\frac{2}{3}$ & 1 & 1 & $\frac{7}{6}$ & $\frac{7}{6}$\\
\hline
\end{tabular}

\section{Conclusion}

We have discussed $N=1$ superconformal theories on the torus and on the cylinder and
derived a generalized Verlinde formula for the fusion coefficients.
For the diagonal theories classified by $(A,A)$ graphs, we have given a lattice
realization of the corresponding superconformal boundary conditions, except for the
trivial fixed point character.
This can be used to study superconformal bulk and boundary flows via TBA \cite{PCA01}.

Using the methods introduced here, the non-diagonal theories can be investigated as 
well.
The corresponding $(A,G)$-type theories, where $G$ is of \ade-type, may be obtained by
constructiong the integrable \ade-lattice models at fusion level $(2,2)$, together with
their superconformal boundary conditions.
Whereas this is a straightforward generalization of the methods presented here (see
also \cite{BP01,MP01}), it is not obvious how to obtain lattice realizations of the 
$(D,A)$ and $(D,E)$ theories.

Focusing on the coset construction, we have given the coset boundary 
conditions corresponding to the branching functions, in the simplest case of the level 
two $\widehat{s\ell}(2)$ coset models corresponding to the $N=1$ superconformal theories.
As for the superconformal fixed point field, this excludes a lattice realization of the
resolved fixed point fields. 
The above methods can be used to obtain integrable and conformal boundary conditions 
for the coset models at fusion level higher than two \cite{KP92} by an obvious 
generalization.
Our claim is that the corresponding coset boundary weights give a complete realization
of coset boundary conditions apart from fixed point fields.

\section*{Acknowledgements}

CR would like to thank the German Research Council (DFG) for financial support.
This work is supported by the Australian Research Council (ARC).

\end{document}